\documentclass[11pt]{article}  
\usepackage{amssymb,amsthm,amsfonts,psfig,epsfig,axodraw,color}
\def\gappeq{\mathrel{\rlap {\raise.5ex\hbox{$>$}} {\lower.5ex\hbox{$\sim$}}}}
\def\lappeq{\mathrel{\rlap{\raise.5ex\hbox{$<$}} {\lower.5ex\hbox{$\sim$}}}}

\def\beq{\begin{equation}} \def\eeq{\end{equation}} 
\def\bea{\begin{eqnarray}} \def\eea{\end{eqnarray}}
\def\bq{\begin{quote}} \def\eq{\end{quote}}

\def\nn{\nonumber}

\def\ti{\tilde}
\def\d{\delta}
\def\meg{\mu \rightarrow e \gamma} \def\tmg{\tau \rightarrow \mu \gamma} \def\teg{\tau\rightarrow e \gamma}
\def\tgb{\tan\beta}

 \def\ie{{i.e.}} \def\eg{{\it e.g.~}}

\begin{document} 

\pagestyle{empty} 
\begin{flushright}  SACLAY-T02/159\end{flushright}  
\vskip 2cm    
\def\thefootnote{\fnsymbol{footnote}}
\begin{center}  
{\large \bf \sc Sleptonarium \\ \vspace{.3 cm} 
(Constraints on the CP and Flavour Pattern of Scalar Lepton Masses)} 
\vspace*{5mm}    \end{center}  
\vspace*{5mm} \noindent  \vskip 0.5cm  
\centerline{\large \bf I. Masina and C. A. Savoy
\protect\footnote{E-mails:  masina@spht.saclay.cea.fr, savoy@spht.saclay.cea.fr.}}
\vskip 1cm\centerline{\em Service de Physique Th\'eorique 
\protect\footnote{Laboratoire de la Direction des Sciences de la Mati\`ere du Commissariat \`a l'\'Energie
 Atomique et Unit\'e de Recherche Associ\'ee au CNRS (URA 2306).} , CEA-Saclay}
\centerline{\em F-91191 Gif-sur-Yvette, France} \vskip2cm   
\centerline{\bf Abstract}  

The constraints on the flavour and CP structure of scalar lepton mass
matrices are systematically collected. The display of the resulting
upper bounds on the lepton -slepton misalignment parameters is designed
for an easy inspection of very large classes of models and the
formula are arranged so as to suggest useful approximations.
Interferences among the different contributions to lepton flavour
violating transitions and lepton electric and magnetic dipole moments
of generic character can either tighten or loose the bounds.  A
combined analysis of all rare leptonic transitions can disentangle the
different contributions to yield hints on several phenomenological
issues. The possible impact of these results on the study of the
slepton misalignment originated in the seesaw mechanism and
grand-unified theories is emphasized since the planned experiments are
getting close to the precision required in such tests.

\vspace*{1cm} \vskip .3cm  \vfill   \eject  
\newpage  
\setcounter{page}{1} 
\pagestyle{plain}


\def\thefootnote{\arabic{footnote}}
\setcounter{footnote}{0}


\section{Introduction}

Neutrino oscillation experiments have established the fact that the
lepton family numbers are violated\cite{oscexp}. Looking upon
the Standard Model (SM) as an effective theory, besides the $d=5$
operator responsible for Majorana neutrino masses, there is a $d=6$
operator where lepton flavour and CP violations could further manifest:  
\beq
\frac{1}{\Lambda^2} {\bar \psi} \sigma^{\mu \nu} (1 + \gamma_5) \psi
F_{\mu \nu} \phi ~~~, \label{bop}
\eeq   
from which lepton flavour violating decays (LFV) $\ell_i \rightarrow
\ell_j \gamma$ and additional contributions to electric and magnetic
dipole moments (EDM, MDM) all potentially arise.  The present upper
limits and the planned future sensitivities to such observables are
displayed in Table \ref{explim}.  If the fundamental theory is well
described by the SM up to very large scales, \eg , up to the gauge
coupling unification scale, then these operators are too much
suppressed to be observed. However, if the new
physics scale is low enough the above processes are potentially
detectable. This is indeed the case for low energy supersymmetric
extensions of the SM where flavour violations would originate from any
misalignment between fermion and sfermion mass eigenstates.
Understanding why all these processes are strongly suppressed is one of
the major problems of low energy supersymmetry, the {\it flavour
problem}, which suggests the presence of a quite small amount of
fermion sfermion misalignment.

In evaluating the above processes, we are thus allowed to use the
so-called mass insertion method \cite{massinsmeth} . This is a particularly
convenient method since, in a model independent way, the tolerated deviation
from alignment is quantified by the upper limits on the mass
insertion $\delta$'s, defined as the small off-diagonal elements in
terms of which sfermion propagators are expanded.  They are of four
types: $\delta^{LL}$, $\delta^{RR}$ $\delta^{RL}$ and $\delta^{LR}$,
according to the chiralities of the sfermions involved.  In principle,
one could test each matrix element of these matrices. Indeed,
searches for the decay $\ell_i \rightarrow \ell_j \gamma$ provide 
bounds on the absolute values of the off diagonal (flavour violating)
$|\delta^{LL}_{ij}|$, $|\delta^{RR}_{ij}|$, $|\delta^{LR}_{ij}|$ and
$|\delta^{RL}_{ij}|$, while measurements of the lepton EDM (MDM), $d_i$
($ a_i$), besides giving constraints on the flavour conserving mass
parameters and their CP violating phases, also provide limits on the
imaginary (real, respectively) part of combinations of flavour
violating $\delta$'s, $ \delta_{ij}^{LL}\delta_{ji}^{LR}$, $
\delta_{ij}^{LR}\delta_{ji}^{RR}$, $\delta_{ij}^{LL}\delta_{ji}^{RR}$
and $ \delta_{ij}^{LR}\delta_{ji}^{LR}$.

Although many authors have addressed the issue of the bounds on these
misalignment parameters in the leptonic sector, the analysis have often
focused more on some particular observables. It is worth 
emphasizing that the study of the combined limits allows to extract 
additional informations, as discussed in this paper. On the
other hand, other more general studies \cite{ggms} only considered the
contribution of a photino inside the loop diagram (this roughly
corresponds to the bino contribution in our work). Other contributions
are often dominant depending on the region of the parameter space.

The aim of our work is to reconsider the limits on scalar lepton masses
in a systematic approach and to lay them out in such a way that one can
easily extrapolate the results from a model to another. In the
following sections, we present a global update of the present limits
and we analyse the impact of the  planned experimental improvements. In
particular, they offer the possibility of learning something about CP
violating phases in the flavour violating elements of the sleptonic
sector. This could have interesting implications from the theoretical
point of view.

The relevant one-loop amplitudes have been exactly written in terms of
the general mass matrix of charginos and neutralinos \cite{cexact},
resulting in quite involved expressions. The results become more
transparent and more suitable for a model independent display in an
approximation where the gaugino-higgsino mixings are also treated as
insertions in the propagators of the charginos and neutralinos inside
the loop \cite{moroi, prs, fms}.  The relevant amplitudes for the
chirality flip processes considered here are then conveniently
classified according to the type of gaugino in the propagator: bino,
higgsino-bino or higgsino-wino. With this additional insertion
approximation, it is not really necessary to fix a particular scenario
to analyse the dependence on the many mass parameters as the relevant
terms become more explicit and one can use simple approximate
expressions to understand the behaviour in some regions of the
supersymmetric parameter space. However, in the figures, it is
convenient to select a reasonable framework so that all the limits on
the $\delta$'s can be easily compared in terms of the same observables.
We consider the mSUGRA scenario and we display the upper bounds on the
$\delta$'s in the $(M_1,m_{\ti e_R})$ plane ($M_1$ and $m_{\ti e_R}$
are the bino and right-slepton masses, respectively) assuming gaugino
and scalar universality at the gauge coupling unification scale and
fixing $\mu$ as required by the radiative electroweak symmetry
breaking.

Deviations from the mSUGRA assumptions can be estimated by means of 
relatively simple analytical expressions. For this sake, the main
contributions are isolated and simple approximations are offered
so to ease the adjustment of the constraints to alternative models.

In our analysis, we pay  attention to the possible interferences among
the amplitudes. Some of them could be introduced in a more artificial
way, \eg by adjusting the phases between the wino and masses, $M_2$ and
$M_1$, or those between $A$ and $\mu$ to suppress the lepton EDM.  We
are more interested in interferences that are generically  present in
the models and would affect the limits on the  $\delta$'s. This happens
for instance for $\delta^{RR}_{ji}$ due to a destructive interference
between the bino and bino-higgsino amplitudes, so that no limit can be
derived in the region where $\mu^2 \approx (m_{\ti e_L}^2 +
m_{\ti e_R}^2)^2/4m_{\ti e_R}^2$, that in mSUGRA translates into
$m_{\tilde e_R} \sim 6 M_1$.  On the contrary, the limits on
$\delta^{LL}_{ji}$ are robust - if $M_2$ and $M_1$ have similar phases
- because of a constructive interference between the chargino and bino
amplitudes. It turns out that in the mSUGRA dark matter favoured region
$m_{\tilde e_R} \approx M_1\, ,$ the bino diagram dominates, while the
chargino one gives the largest contribution above the sector $2 <
m_{\tilde e_R}/ M_1 < 3\, ,$ where they have comparable strength.  More
generally, the bino takes over the chargino around $ |\mu |^2 /m_{\ti
e_L}^2 \sim 1-10 ~,$ depending on the model.  As a consequence the
limits on $\delta^{LL}_{ji}$ uniformly decrease along any direction of
the $(M_1,m_{\tilde e_R})$ plane.

The present limits on $\mu \rightarrow e \gamma$ and $d_e$ already
provide interesting constraints on the related $\delta$'s. A sensitivity
could be reached in future experiments on $\tau \rightarrow \mu \gamma$
and $d_\mu$  that would allow to test the values of the $\delta$  at
the level of the radiative effects, as predicted, for instance, in the
see-saw context \cite{bormas}.

Another issue is the origin of the CP violating phases in the lepton
EDM. Unless the sparticle masses are considerably increased, the phases
in the diagonal elements of the slepton masses (in the lepton flavour
basis), involving the parameters $\mu$ and $A_i$ of supersymmetric
models, have to be quite small, which is the so-called supersymmetric
CP problem. Thus, one would like to establish if they could arise
instead from flavour violating contributions with $O(1)$ phases,
analogously to the large CP violating phase in the CKM matrix - in
spite of the different origin of the mass misalignments. The bounds on
the $\delta$'s from LFV decay experiments, set limits to the sole LFV
contributions to EDM.  Those obtained from the searches for $\tau
\rightarrow \mu \gamma$ give limits on the LFV part of $d_{\mu}$,
namely, that coming from LFV mass insertions \cite{fms}.  The present
limits on the appropriate $\delta$'s still allow for a much larger
$d_\mu$ from LFV than what is expected from the lepton flavour conserving (LFC)
ones on the basis of the present limits on $d_e$ and the mass scaling
rule \cite{fms, romstr}.  Indeed, the LFV contributions to EDM are
likely to strongly violate this rule.  

The paper is organized as follows. In Section 2 we define the
insertion approximations and the general expressions. The limits
on the different $\delta$ matrix elements obtained from the
present and future experimental bounds on lepton flavour violating 
leptonic decays are displayed and commented upon in Section 3.
In Section 4 we separate the analysis of the lepton flavour
conserving and violating contributions to the lepton MDM and EDM, 
and in Section 5 we discuss the possibility of isolating the respective sources
of CP violations by combining the data on EDM and LFV decays.
In Section 6 we draw our conclusions and in the Appendix we exhibit the analytic expressions 
of the various contributions to the processes discussed in this paper
in the insertion approximation as well as some useful approximations.

\begin{table}[!ht]
\begin{center}
\begin{tabular}{|c||c|c|} 
\hline & & \\
& present & planned \\ & &  \\ \hline  \hline &  & \\ 
$d_e$ & $ < 1.5~ 10^{-27} $ e cm \cite{deexpp} & $< 10^{-29 (-32)}$ e cm \cite{deexpf}(\cite{lam}) \\ &  &  \\  \hline  &  & \\ 
$d_\mu$ &  $< 10^{-18} $ e cm \cite{dmuexpp}   &  $ < 10^{-24 (-26)}$ e cm  \cite{dmuexpf}  (\cite{dmuexpff}) \\  & & \\  \hline   &  & \\
$d_\tau$ & $< 3~ 10^{-16}$ e cm \cite{PDB} &   \\ & &  \\  \hline & & \\
BR($\mu \rightarrow e \gamma$) & $ < 1.2~ 10^{-11} $  \cite{PDB} & $< 10^{-14}$  \cite{megexpf} \\ &  &  \\  \hline  &  & \\ 
BR($\tau \rightarrow \mu \gamma$) &  $<1.1~ 10^{-6} $  \cite{PDB}   &  $ < 10^{-9}$(?)  \cite{tmgexpf}  \\  && \\  \hline   &  & \\
BR($\tau \rightarrow e \gamma $) &  $<2.7~ 10^{-6} $  \cite{PDB}   &    \\ & &  \\  \hline 
\end{tabular}
\end{center}
\caption{Present experimental limits and planned sensitivities to lepton electric dipole moments and flavour
violating decays.}
\label{explim}
\end{table}


\section{Framework}\label{frame}

In this paper we are concerned with the supersymmetric contributions to the electromagnetic dipole
transitions of leptons, induced by 
\beq \
\frac{e}{4}~M_{ij}~{\bar\psi}_j \sigma^{\mu \nu}\psi _i F_{\mu\nu} ~~~,
\label{M} 
\eeq 
\noindent from which the FC $(i=j)$ EDM transitions and additional
contributions to MDM ones as well as lepton flavour violating $(i \neq
j)$ transitions, all potentially arise.  The contributions of broken
supersymmetry to $M_{ij} ~,$ because of its dimension, must have a
factor $m_{susy}^{-1}\, ,$ of the scale of the soft supersymmetry
breaking mass parameters, $m_{susy}\, .$ However, the transitions
induced by (\ref{M}) have a chirality flip character, hence an
additional factor $m_j / m_{susy}$ will be present. Indeed, the L-R
character of the operator in (\ref{M}) requires at least one Higgs
v.e.v., $v ~,$ factor,  by isospin and hypercharge conservation, as
already exhibited by the basic gauge invariant operator (\ref{bop}).
This can appear in the loop only through either a $ I= 1/2 $ mass terms
between L and R sfermions or between  gauginos and higgsinos. The
chirality flip can only be provided by a Yukawa coupling of either the
higgsino at a vertex or by the chirality flip sfermion masses, also
related to Higgs Yukawa couplings. Hence, a $m_j / m_{susy}^2$ factor
can always be factorized in (\ref{M}). Of course, the variety of the
soft masses which, together with the $\mu$ term, enter the exact
expression of $M_{ij}$ requires a more precise calculation, also
justified by the presence of several interfering amplitudes.


\subsection{Insertion approximations} \label{approximations}

The calculation of these amplitudes in flavour space encompasses a large
number of soft mass parameters from the chargino-higgsino sector as
well as the mass matrix for all left- and right-handed sleptons and
left-handed sneutrinos.  However, this complexity is conveniently
reduced without spoiling the physical results by two kinds of so-called
insertion approximations for the sparticles inside the loop that we now
specify.

\subsubsection{Chargino-neutralino branch.}\label{insertion1}

The MDM/EDM one-loop amplitudes have been fully calculated in the
literature \cite{amusus1, amusus2, dsus1, dsus2, prs} and have also been displayed
\cite{moroi, fms} in the insertion approximation as a development in
powers of $M_Z / m_{susy}$ of the chargino and neutralino propagators.
This approximation is very satisfactory in most of the parameter space
\cite{moroi, prs} and it simplifies the analysis of the dependence of
the results on the soft masses. One has: the non-flip components in the
propagator of the gauginos, $\tilde{B}~,$ with mass $M_1~,$ and
$\tilde{W}~,$ with mass $M_2~,$ and of the higgsinos $\tilde{H}$ with
mass $\mu $ (the contributions of the latter are  smaller by a factor $
\sim m_i / M_Z ~) ~;$ and the flip components $\tilde{B} - \tilde{H} ~,
~\tilde{W} - \tilde{H} ~, $ proportional to $v ~.$

\begin{figure}[!ht]  
\begin{center}  

\begin{picture}(500,100)(0,0)  

\Line( 0,15)(120,15)
\Text( 5,5)[]{${\ell_L}$} \Text(115,5)[]{${\ell_R}$}
\Text(60,5)[]{${\tilde B}$}
\DashCArc(60,15)(40,0,180){5}
\Text(10,35)[]{$ {\tilde e}_{L}$} \Text(110,35)[]{$ {\tilde e}_{R}$}
\Vertex(60,55){3}
\Photon(95,55)(135,90){3}{4}

\Line( 180,15)(300,15)
\Text( 185,5)[]{${\ell_L}$} \Text(295,5)[]{${\ell_R}$}
\Text(220,5)[]{${\tilde H}^0$} \Text(260,5)[]{${\tilde B}$}
\Text(220,-10)[]{$({\tilde B})$} \Text(260,-10)[]{$({\tilde H}^0)$}
\DashCArc(240,15)(40,0,180){5}
\Text(240,80)[]{$ {\tilde e}_{R}$} \Text(240,65)[]{$( {\tilde e}_{L} )$}
\Vertex(240,15){3}
\Photon(260,50)(305,90){3}{4}

\Line( 360,15)(480,15)
\Text( 365,5)[]{${\ell_L}$} \Text(475,5)[]{${\ell_R}$}
\Text(400,5)[]{${\tilde W}^0$} \Text(440,5)[]{${\tilde H}^0$}
\Text(400,-10)[]{$({\tilde W}^\pm)$} \Text(440,-10)[]{$({\tilde H}^\pm)$}
\DashCArc(420,15)(40,0,180){5}
\Text(420,80)[]{${\tilde e}_{L}$}  \Text(420,65)[]{$ ({\tilde \nu})$}
\Vertex(420,15){3}
\Photon(440,50)(475,90){3}{4} \Photon(440,20)(475,60){3}{4}
\Text(437,23)[]{$($} \Text(481,58)[]{$)$}

\SetColor{Blue}
\Text(60,-40)[]{ \Large $(\tilde B)$}
\Text(240,-40)[]{ \Large $(\tilde H^0 \tilde B)$}
\Text(420,-40)[]{\Large $(\tilde W \tilde H)$}

\end{picture}

\end{center} 
\end{figure}

\vskip 1cm
\noindent {\large $~~~~~~~~~~~\underbrace{~~~~~~~~~~~~~~~~~~~~~~~~~~~~~~~~~~~~~~~~~~~~~~~~~~~~~~~~~~}
~~~~~~~~~~~~~~~~~~~~~~~~~~~\underbrace{~~~~~~~~~~~~~~~~~~~}$}\\
{\center {\LARGE  ~~~~~~~~~~~~~~~~~~~~~~~~~~$U(1)$~~~~~~~~~~~~~~~~~~~~~~~~~~~~~~~~~~~~~~~~~~~~$SU(2)$}}
\vskip 1cm

The approximation consists in keeping at most one flip insertion,
namely, only the terms of $O(v/ M_i ~, v/ \mu)~.$ Actually, it can be
considered as the lowest term in the development in powers of $ v^2 /
m_{susy}^2 $, which is precisely the amount of fine-tuning in the MSSM.
This gives a theoretical motivation for this approximation, which is
corroborated by the numerical checks.

In this insertion approximation, the expressions for MDM, EDM and LFV
decays are all obtained from the three chirality-flipping amplitudes
associated to the Feynman diagrams displayed above, where the $v$
insertion is also shown. Their expressions, as given in the Appendix,
allow for a simple factorization that suggest useful approximations in
most of the cases.  As for CP phases in the chargino-neutralino
sector, they are defined so that $M_1$ is real while $\mu$ and $M_2$
remain complex in general.


\subsubsection{Slepton branch: the $\delta '$s.}\label{insertion2}

It is convenient to work in the basis where lepton flavour is defined
so that the charged lepton masses, the Yukawa couplings and the
gauge couplings are flavour diagonal. In general, it is not necessarily
so for the slepton mass matrix in the same basis, where the
non-diagonal entries induce the FV effects in $M_{ij} ~,$ as they
measure the misalignment between the lepton and slepton physical
states.  Because of the severe experimental constraints on the LFV
transitions and on the CP violating EDM's, it is well justified to
develop the slepton propagators around the diagonal terms so defining
an approximation where the non-diagonal terms appear as insertions. On
the other hand, consistency with the previous approximation in $ v^2 /
m_{susy}^2$ means that the mass splitting inside an isospin doublet, $
\propto M_Z^2 ~,$ should be neglected, \ie , the sneutrinos and charged
sleptons are mass degenerate. More pragmatically, since sneutrinos
appear in the chargino diagram which dominates in several cases, this
FC mass splitting only affects the results in the limit where all the
soft terms are small, generically disfavoured by present experiments.

Along these lines, we shall adopt here the usual convention 
for the slepton mass matrix in the basis where the lepton mass matrix
$m_{\ell}$ is diagonal: 
\beq
\left( \begin{array}{cc} {\tilde \ell}_L^\dagger & 
{\tilde \ell}_R^\dagger \end{array} \right) \ \ 
\left( \begin{array}{cc} m_L^2 (\mathbb{I} + \delta ^{LL} ) &
(A^* - \mu \tan\beta) m_{\ell} + m_L m_R\delta ^{LR} \cr
(A - \mu^* \tan\beta) m_{\ell} + m_L m_R\delta ^{LR~\dagger} &
m_R^2 ( \mathbb{I}+ \delta ^{RR} )
\end{array} \right)
\ \ \left( \begin{array}{c} {\tilde \ell}_L  \cr {\tilde \ell}_R 
\end{array} \right) \label{sleptonm2}
\eeq 
\noindent where $m_L ~, m_R ~,$ are respectively the average real
masses of the L and R sleptons and $A \sim O(m_{susy})$ is the diagonal
mass matrix of the $A-$term.  The deviations from this universal mass
matrix are all gathered in the $\delta$ matrices, which contain 30 real
parameters (including 12 phases). Those in $\delta ^{LR}$ are expected
to be proportional to the $m_{\ell}$ eigenvalues with $O(m_{susy})$
coefficients, while, by their own nature, $m_L^2 ~, m_R^2 \sim
O(m_{susy}^2)$.  But, from the experimental constraints on LFV and EDM,
important suppression factors with respect to $m_{susy}$ are expected
in basically all of them to solve the so-called supersymmetric flavour
problem.  Our aim here is to quantify these requirements. Notice that
the diagonal entries in the $\delta$'s (in the present basis) could be
large, but in many cases this can be afforded for by some obvious
corrections in the final results, as we shall discuss.  The insertion
approximation now corresponds to a development of the propagators
around the diagonal with the average slepton masses, $m_L^2$ and
$m_R^2 ~.$

The complete expression for the amplitudes  are given in
the Appendix, in (\ref{all})\footnote{We have checked that the
insertion approximation gives essentially the same results as the exact
expressions in ref. \cite{cexact} with the exception of some very low
energy region.}. These amplitudes are indicated by the
lower indices: $B$ for the pure $\tilde{B}$ diagram, $L$ and $R$ for
the $\tilde{B}\, -\, \tilde{H}^0$ one with $L$ and $R$ sleptons,
respectively, and $2$ for the $\tilde{W}\, -\, \tilde{H}$ ($SU(2)$)
one. In this development, the non-diagonal insertions
generate LFV transitions, but also contribute at a higher level of the
development to the FC transitions, in particular by generating EDM
phases.  Therefore we expand beyond the first term and define a
multiple insertion approximation. Thus, upper bounds can be used to
constrain not only single matrix elements of the $\delta$'s, but also
some of their products.


\subsection{Order of magnitude of the amplitudes} \label{estimates}

The advantage of the mass insertion approach is precisely that it 
disentangles the various contributions which depend on different sets of 
parameters. Furthermore, as shown in the Appendix, each one can be 
factorized in two terms, as shown in (\ref{Iapp}). 

The first important observation is that all contributions  get a factor
of $\tan\beta$. Even the pure $\tilde{B}$ one has a factor $( \mu^* \tan
\beta - {A} )$ in the diagonal terms. This is the well known fact that
MDM and EDM are roughly proportional to $\tan\beta$ for large $\tan\beta$.
For this reason, it is convenient to introduce the complex parameters 
\beq
\eta_i = 1- \frac{A_i}{\mu^* \tgb} \, .
\eeq

Let us identify the main factors in the contributions of the different
Feynman graphs, in the most plausible scenarios where $|\mu |
\gg M_i$ is required, as well as $M_1 < m_L ~, m_R ~.$ We  denote
$M^{(2)}$ the contribution from the $SU(2)$ graph, $M^{(B)}$ the pure
$\tilde{B}$ one, $M^{(L)}$ and $M^{(R)}$ those from the other two
$U(1)$ graphs.  Then the main contributions to, \eg , the diagonal
terms can  be approximated as follows (in this particular example, we
make the obvious replacement of the average slepton masses by the
relevant eigenvalue):
\bea
M^{(2)}_i & \approx & \frac{\alpha M_2^* m_i \tgb}{4\pi\mu m_{L_i}^2 
\sin^2 \theta_W}\ g_2 \left( x'_{L_i}\right) \label{M^2} \\
M^{(L)}_i & \approx & \frac{\alpha M_1 m_i \tgb}{8\pi\mu m_{L_i}^2 
\cos^2 \theta_W}\ g_1 \left( x_{L_i}\right)\label{M^L} \\
M^{(R)}_i & \approx & -\frac{\alpha M_1 m_i \tgb}{4\pi\mu m_{R_i}^2 
\cos^2 \theta_W}\ g_1 \left(x_{R_i}\right)\label{M^R} \\
M^{(B)}_i & \approx & \eta_i \frac{|\mu |^2}{m_{R_i}^2 - m_{L_i}^2} 
\left( 2M^{(L)}_i + M^{(R)}_i \right) \label{M^B}
\eea
\noindent where we have introduced the notations:
\beq
x_L= \frac{M_1^2}{m_L^2}
\qquad x_R= \frac{M_1^2}{m_R^2} \qquad x'_L= \frac{|M_2|^2}{m_L^2} 
= \frac{|M_2|^2}{M_1^2}x_L
\eeq 
\noindent and $g_1$ and $g_2$ are displayed in the Appendix and in
fig. \ref{Ffunc_ghk}.  Roughly, $g_1 \sim O(1)$ for  $M_1 < m_L ~, m_R
~,$ and $g_1\sim m_{L(R)}^2 / M_1^2$ in the opposite situation.
Instead, $g_2$ feels the chargino mass singularity due to collinear
photons, with a logarithmic singularity for  $|M_2|^2 \rightarrow 0 ~,$
and $g_2 \sim m_{L}^2 /( 2|M_2|^2)$ for $m_{L}^2 \ll |M_2|^2$.

From the symmetries of the equations in (\ref{Iapp}) it is easy to
adapt (\ref{M^2}) -- (\ref{M^B}) to other configurations of the
parameter space. A rule of thumb for a rough evaluation: to rescale 
these results from $M_1 < m_R ~(m_L)\, ,$ to $M_1 > m_R ~(m_L)\, ,$
multiply the amplitudes $M^{(R)}$ ($M^{(L)}$) as given in
these expressions by $x_R^{-1}$ ($x_L^{-1} \, ,$ respectively), and
analogously for $|M_2| > m_L$ and $M^{(2)}\,$ (this is not as good
because of the mass singularity though). To adjust from $|\mu |\gg M_i$
to $|\mu |\ll M_i \, ,$ one just exchanges $|\mu| \leftrightarrow M_1\,
,$ in $M^{(R)}$ and $M^{(L)}$, $|\mu|\leftrightarrow |M_2|\, ,$ in
$M^{(2)}$ but not in $M^{(B)}\,$.

The other terms in the insertion expansion, namely, the
coefficients of $\delta$'s, are similar, but slightly suppressed by one
or two more propagators. They will be discussed more in detail in the
next sections. But the general trend is given by (\ref{M^2}) --
(\ref{M^B}).

A generic remark is in order. All the contributions in (\ref{M^2}) --
(\ref{M^B}) have the same sign but $M^{(R)}$, due to the negative
hypercharge of the L sleptons, at least for $ |\mu | \tan\beta > |A|
~,$ and for $M_2$ in phase with $M_1$. Its destructive interference
with the others has some interesting consequences to be discussed later
on. 

Actually, many of the results in the figures can be understood from
these simple approximations. More importantly, in spite of these
figures being displayed within the mSUGRA constrained parameter space,
the results can be easily extrapolated to other models from the
approximated expressions in the Appendix and their adaptations.
In most cases one can identify a dominant contribution and a
corresponding approximation, then the relation with the plots in
the figures, just as we do for some examples below.


\subsection{mSUGRA spectrum} \label{spectrum}

The dependence of the results on the parameters $(~ \mu ~, A
~,M_1 ~, M_2 ~, m_L ~, m_R~ )$ is already transparent in the insertion
approximation. However, it is convenient to display all the bounds on the
various $\delta $'s in the plane of two physical observables, because
this allows to easily discuss, compare and generalize the upper bounds.
In the following pictures, we have chosen to display all the bounds on
the various $\delta $'s in the plane $(m_R, M_1).$ One of the
advantages of this choice is that general cosmological constraints
(see for instance \cite{cosmo}) 
are more easily discussed in this plane.  To reduce the
number of free parameters to just those, we adopted the mSUGRA spectrum
and we provide some rules which allow to consider much more general
scenarios.

Let us recall the constraints arising in mSUGRA, where the universality
assumption reduces the parameters which are then defined close to the
Planck scale as a general scalar mass, $m_0 ~,$ an overall gaugino
mass, $M_{1/2} \, ,$ and a universal $A_0$, together with the low
energy ones, $\mu$ and $\tan\beta~.$ At the low energy scale, say $v$,
after the RGE running, the parameters $M_1 ~, M_2 ~, m_L ~, m_R~,$ are
obtained as follows:
\bea
M_i (v) &=& \frac{\alpha_i(v)}{\alpha_i(M_U)} \, M_i (M_U) 
\qquad (i=1,2,3) \nn \\
m^2_R (v)&=& m^2_R (M_U) + 0.15 M_1^2 (M_U) \nn \\
m^2_L (v) &=& m^2_L (M_U) + 0.51 M_2^2 (M_U) + 0.04 M_1^2 (M_U) \label{masses}
\eea
\noindent where $M_U$ is the unification scale and 
the contribution of Yukawa couplings is included in the $\delta $'s.
The mSUGRA constraints are fulfilled by putting: 
$M_1 (M_U) = M_2 (M_U) = M_{1/2}$ and $m^2_R (M_U) = m^2_L (M_U) = m_0^2 \, .$

A very important constraint in mSUGRA comes from the radiative
electroweak breaking condition that requires a fine-tuned $\mu$ in
order to fulfill the minimum condition:
\beq
|\mu |^2=-\frac{m_Z^2}{2} + \frac{1+0.5 \tan^2 \beta}{\tan^2 \beta - 1} m_0^2 
+ \frac{1+3.5 \tan^2\beta}{\tan^2 \beta - 1} M_{1/2}^2 ~~~. 
\label{finetuning}
\eeq
\noindent The most important term in this fine-tuning comes
from gluino loop contributions which is of course reduced if the
gaugino universality is given up and light gluinos are assumed. This
has an impact in the results, and in some of the approximations that
assume  $|\mu|^2 > M_1^2$. The term in $m_0^2$ is more involved in the
general case, the relation (\ref{finetuning}) for $\mu$ is modified if
one relaxes the mass universality between Higgses and matter fermions
at $M_U \, ,$ but $m_0^2$ has a smaller coefficient.

Finally, we replace the mSUGRA variables $m_0$ and $M_{1/2}$ by the
more physical masses $m^2_R (v)$ and $M_1 (v)$, and we suppress the
value $v$ from everywhere. Then $M_2$, $\mu^2$ (for $\tan^2 \beta \gg 1$) 
and $m^2_L$ are fixed as follows:
\beq
M_2 \approx 2 M_1 \qquad |\mu |^2 \approx 0.5 m^2_R + 20 M_1^2 \qquad
m^2_L \approx m^2_R  + 2.5 M_1^2  ~~~. 
\label{mSUGRArelation}
\eeq
These relations are assumed in the figures.
Therefore the two typical regions in the $(m_R, M_1)$ (semi-) plane are $m_R \approx M_1 $ which
is preferred by the mSUGRA dark matter solution, and 
$m_R \gg M_1 $ where $m_L \approx m_R\, .$ 
As discussed later on, notice that in the first
region the bino provides not just the dark matter of the universe,
but also the main contribution to many processes.

In the figures of the next sections the plots indicate the excluded
regions as follows. The light grey region is unphysical, since
$m_0^2<0$ there. The dark grey region is excluded
because the LSP would be the right-slepton instead of $\tilde B$.
The line $m_0^2=0$ is the boundary between them.


\section{LFV decays $\ell_i \rightarrow \ell_j \gamma$}\label{sec:lfv}

The non-observation of the rare decay $\ell_i \rightarrow \ell_j \gamma$
provides upper limits on the absolute value of the four flavour
violating $\delta_{ij}$'s.  Indeed, the corresponding branching ratio
receives the following dominant contributions,
\bea
&& \!{\rm BR}(\ell_i \rightarrow \ell_j \gamma) = 3.4\times 10^{-4}~
{\rm BR}(\ell_i 
\rightarrow \ell_j \bar\nu_j \nu_i)~ 
\frac{M_W^4 M_1^2 \tan^2\beta}{|\mu |^2}
\times \nn \\
& & \! \left\{
\left|\ \delta^{LL}_{ji} ( \eta_i^* I'_{B,L} + \frac{1}{2} I'_L + I'_2 )  
+ \delta^{LR}_{ji}\frac{m_R m_L}{\mu m_i \tan\beta}I_B \right|^2 + 
\left|\delta^{RR}_{ji}( \eta_i I'_{B,R} - I'_R ) + 
\delta^{RL}_{ji}\frac{m_R m_L}{\mu^* m_i \tan\beta} I_B \right|^2
\right\} \label{BRlfv}
\eea
\noindent where the integrals $I$'s are defined in the Appendix,
together with some useful approximations, and they are all positive in
the physically relevant region for the masses, but for $I'_2$ which has
the sign of $M_2 /M_1\, .$ We take advantage of the lepton mass hierarchy
to neglect terms $O(m_j/m_i)$. For relatively large $\tgb $ the coefficient
$\eta_i$ is positive, at least in more usual models, which we assume
unless stated otherwise. The interferences between the different
contributions can influence and even spoil the limits on $\delta$'s as
we now turn to discuss.  Assuming that no accidental cancellations occur
between the flavour structure of the $\delta_{ij}$'s and the
dependence of the integrals on the mass parameters, one puts bounds on
each $\delta$. For instance, for $\tmg$, we obtain limits on different $|\d_{23}|$'s  
from the expression:
\bea
\left(  \left|   \delta^{LL}_{23} ( \eta_3^* I'_{B,L} + \frac{1}{2} I'_L + I'_2 ) \right| 
       + \left| \delta^{LR}_{23}\frac{m_R m_L}{\mu m_\tau \tan\beta}I_B \right| \right)^2 + 
\left(  \left|   \delta^{RR}_{23} ( \eta_3 I'_{B,R} - I'_R ) \frac{}{} \right| 
+ \left| \delta^{RL}_{23} \frac{m_R m_L}{\mu^* m_\tau \tan\beta} I_B \right| \right)^2
\nn\\
\le  1.7 \times 10^4  \frac{|\mu |^2}{M_W^4 M_1^2 \tan^2\beta} 
{\rm BR}(\tau \rightarrow \mu \gamma)
~~~~~~~~~~~~~~~~~~~~~~~~
\label{lfvbis}
\eea

In the following we analyse the dependences of the
limits on these $\delta$'s in mSUGRA and also in less constrained
frameworks. Because the overall phases are not relevant, we omit them
in the equations of this section for simplicity.


\subsection{Limits on $\d^{LL}$}\label{sec.dLL}

\begin{figure}[!t]
\centerline{\psfig{file=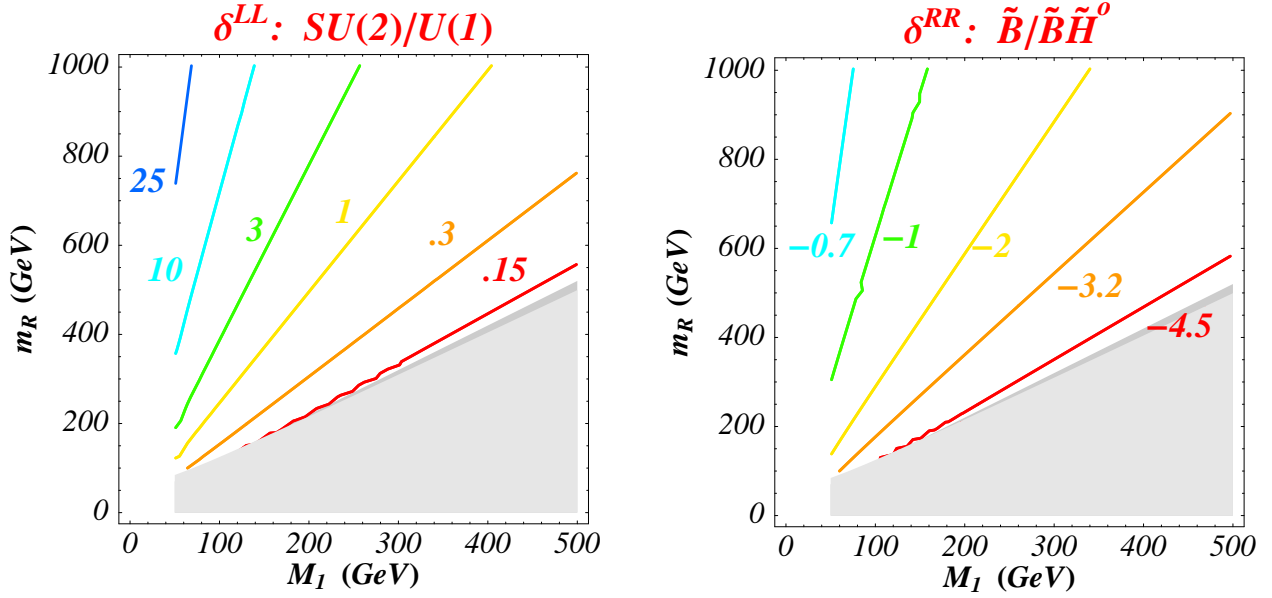,width=1\textwidth}}
\caption{Ratio between the $SU(2)$ and $U(1)$ contributions in the coefficient of $\d^{LL}$
and between the $\ti B$ and $\ti H \ti B$ contributions in the coefficient of $\d^{RR}$.}
\label{rappamp}
\end{figure}

The coefficient of $\d^{LL}$ receives both $U(1)$ and $SU(2)$-type
contributions, respectively from $\tilde B$ and $\tilde B - \tilde
H^0$ exchange and from $\tilde W -\tilde H$ exchange.  Before
discussing the dependences in general supersymmetric scenarios, let us
firstly focus on mSUGRA, where no cancellation between the $U(1)$
and $SU(2)$ amplitudes can arise because they have the same sign.
Their ratio, $I'_2/(\eta_i I'_{B,L} + \frac{1}{2} I'_L)$, is displayed
in fig. \ref{rappamp}. It can be seen from the figure that if $m_R <(>)
2.5 M_1$, then the most important contribution in determining the bound
on $\d^{LL}$ is the $U(1)$ ($SU(2),$ respectively) one. This is quite obvious from
the approximation given in the Appendix for the three contributions for
$|\mu^2 | \gg |M_1^2 |$ as in the mSUGRA case:
\beq
I'_2+I'_{B,L}+\frac{1}{2} I'_L \approx \frac{ M_2\cot ^2\theta_W}
{M_1 m_L^2}\, h_2 (x'_L)+\frac{\mu ^2}{2\bar{m} ^4}\, 
\left( h_1(\bar{x}) +k_1(\bar{x})\right) 
+ \frac{1}{2m_L ^2} \, h_1(x_L) \nn \\  \label{approxdLL}
\eeq
\noindent With the gaugino universality relation $M_2 \approx 2M_1\, ,$
and neglecting the $k_1$ term for simplicity, the region where the
$U(1)$ and the $SU(2)$ contributions are commensurate corresponds to
the condition $h_2(4x_L) /h_1(x_L) \approx (\mu^2 +m_L^2) /13m_L^2\, .$
Then, from the mSUGRA relations (\ref{mSUGRArelation}) one finds $m_R^2
\approx 7M_1^2 \, ,$ close to the numerical result. Notice that the
ratio between the two $U(1)$ terms, in the region where they are
relevant, is $2I'_{B,L}/ I'_L \approx \mu^2 / m_L^2\, .$ In the dark
matter favoured region, $m_R^2 \approx M_1^2\, ,$ and only there, the
pure $\tilde B$ contribution affords for most of the overall rate (up
to 85\%). In the opposite situation, $m_R^2 \gg M_1^2\, , \mu^2 \sim
O(m_L^2)\, ,$ $I'_2/(\eta_i I'_{B,L} + \frac{1}{2} I'_L) \rightarrow
6.8 h_2(x_L) \, ,$ which increases with the mass singularity and yields
a good approximation in the chargino dominance sector \footnote{ 
The $SU(2)$ contribution can be identified with the chargino one,
since the latter is always much bigger than the corresponding neutralino 
contribution.}.

\begin{figure}[!t]
\centerline{\psfig{file=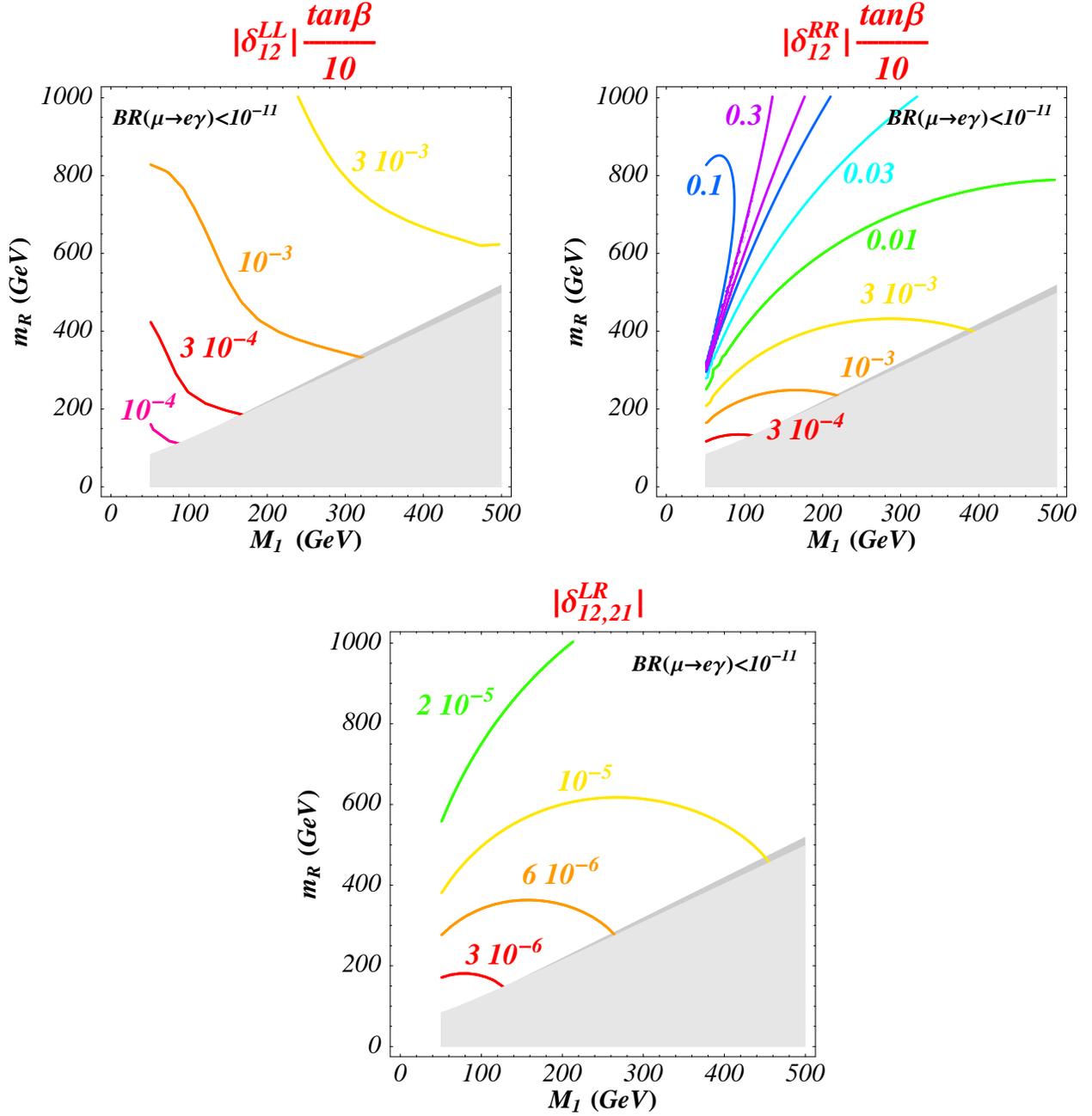,width=1\textwidth}}
\caption{ Upper limits on $\d_{12}$'s in mSUGRA.}
\label{4d12}
\end{figure}

If one relaxes even more the mSUGRA constraints, it becomes legitimate to
ask if one can escape the LFV limits on  $\d^{LL}$'s or, conversely,
how model independent are, \eg , the more stringent limits on
$\d^{LL}_{12}\, .$ The only possibility is to play with violations of
gaugino mass universality. 
For instance, by reducing $|\mu |$ (\ie , the gluino mass, in current models) 
the $\tilde B - \tilde H^0$ and $\tilde W -\tilde H$ contributions 
increase as $|\mu |^{-2}$, while the $\tilde B$ contribution is $|\mu |$ 
independent. Therefore one needs opposite phases for $M_2$
and $M_1\, ,$ in which case the $\d^{LL}$'s would remain unconstrained
inside a relatively narrow sector of the $(m_R, M_1)$ semi-plane.

\begin{figure}[!t]
\centerline{\psfig{file=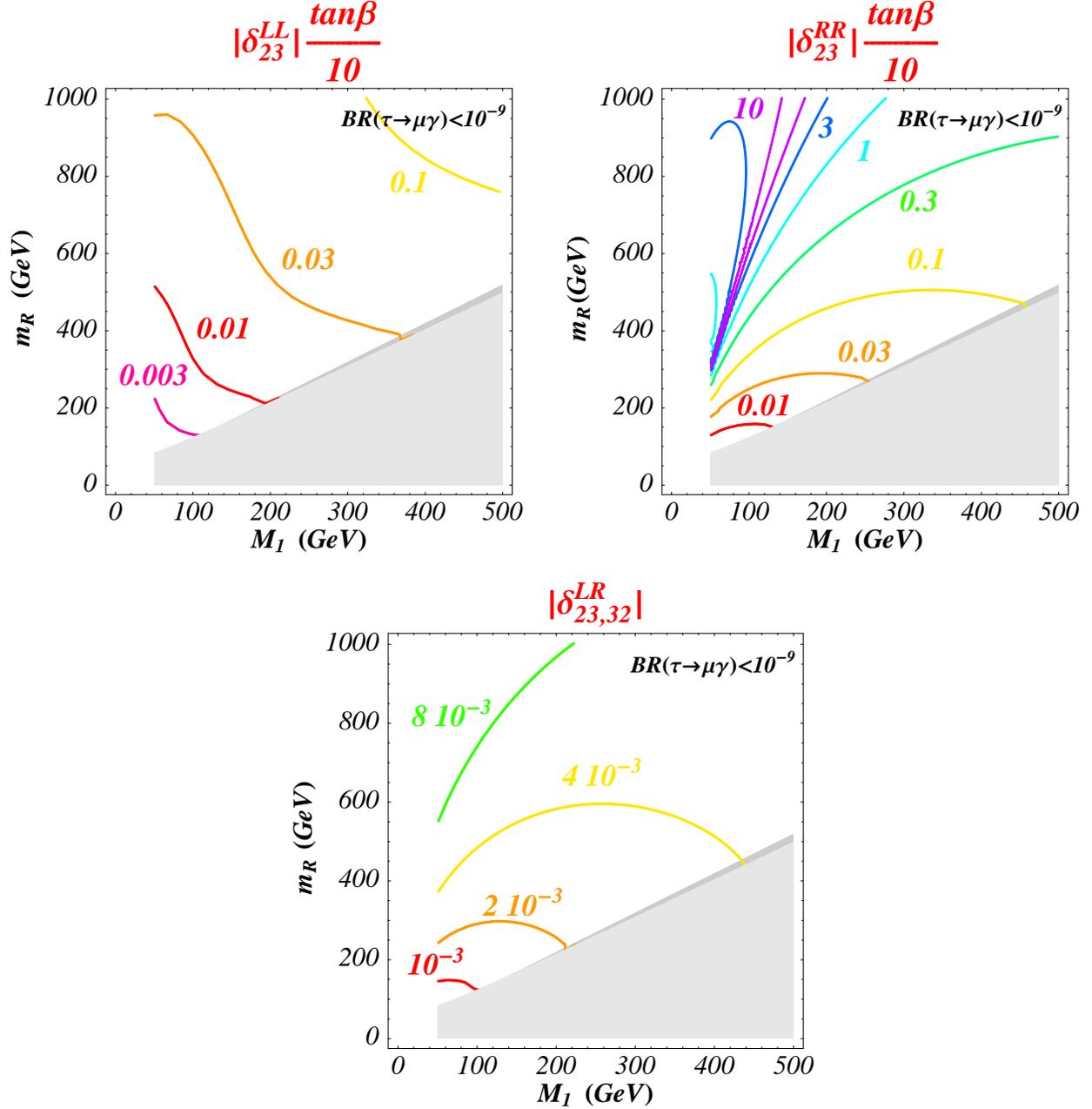,width=1\textwidth}}
\caption{Upper limits on $\d_{23}$'s in mSUGRA.}
\label{4d23}
\end{figure}

In figs. \ref{4d12} and \ref{4d23} we show the global bounds on
$\d^{LL}_{12} \tgb/10$ and $\d^{LL}_{23} \tgb/10$ that follow from
$BR(\meg)< 10^{-11}$ and  $BR(\tmg)< 10^{-9}$ respectively. The former
corresponds to the present bound (see Table \ref{explim}). Since the
branching ratio is quadratic in the $\d$'s, the planned improvement by
three orders of magnitude on $BR(\meg)$ would strengthen the limit by a
factor of 30. 
\footnote{
Actually, $\mu-e$ conversion in nuclei (see for instance \cite{kko} and 
references therein) also gives a bound \cite{meconv} on $\meg$. 
There is a proposal to achieve a precision at the level of $10^{-16}$ in
$\mu-e$ conversion \cite{meconvf}, which would imply a really strong 
limit on the flavour dependence of the left-handed sleptons, comparable
to $BR(\meg)$ at the level of $10^{-14}$.} 
Notice that the bound decreases quite uniformly as $m_{susy}^{-2}$ thanks 
to the positive interference among the three contributions.

Instead, $BR(\tmg)< 10^{-9}$ is more like a somewhat optimistic
prospect. Actually, the present limit on $\d^{LL}_{23} \tgb/10$ is
larger by a factor of 30. Thus, depending on $\tgb$, $\d^{LL}_{23}$ is
still poorly constrained in most of the mSUGRA parameter space. Here
and in the following, we display the bounds corresponding to the
planned sensitivity in order to stress the relevance of such an
experimental progress. Indeed, at this level, the precision in
$\d^{LL}_{23}$ is at the level of the radiative corrections induced by
the seesaw mechanism, providing a unique test for the origin of the
atmospheric neutrino oscillations 
\cite{lms1}. 
As for $\teg$, the limit on $\d^{LL}_{13}$ is such that $\d^{LL}_{13}/\d^{LL}_{23}=
(BR(\teg)/BR(\meg))^{1/2}$; thus, the present bound on $\d^{LL}_{13}
\tgb/10$ is worse by a factor of about 50 with respect to fig. \ref{4d23}.

\begin{figure}[!ht]
\centerline{\psfig{file=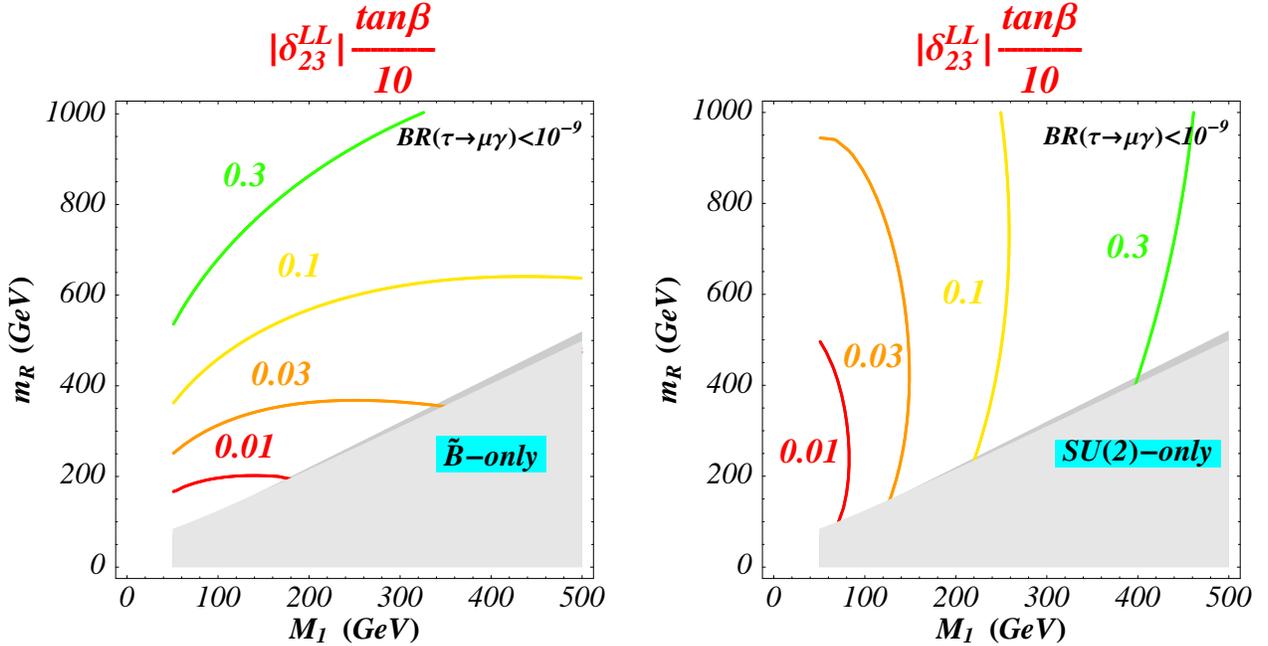,width=1\textwidth}}
\caption{Limits on $\d_{23}^{LL} \tan\beta/10$ obtained by considering only the $\ti B$ and
the $SU(2)$ amplitude respectively.}
\label{d23cn}
\end{figure}

To stress the importance of considering both the $\ti B$ and the $\ti H
-\ti W$ contributions, the bound on $\d^{LL}_{23}$ is shown in fig.
\ref{d23cn} by taking into account only the $\tilde B$ or the $\tilde W
-\tilde H$ one (analogous considerations apply for $\mu \rightarrow e
\gamma$). The alternate dominance of the $\tilde B$ and the chargino
contributions according to the value of $M_1^2/m_R^2$ is again quite evident.
Some of the previous analysis have only considered the chargino part of
the $SU(2)$ contribution \cite{ell, lms1}, or only the
$\tilde B$ contribution \footnote{Actually, the photino was considered
in Ref. \cite{ggms} , but it corresponds to the $\tilde B$, with
$\alpha _1 \rightarrow 2 \alpha_{\rm em}$ for the same gaugino
masses.}.

The chargino dominance when $|\mu |$ is smaller than the slepton masses
is an interesting feature since the other mass misalignments leading to
LFV, $\d^{RR}$ and $\d^{LR}$ are not as effective for this mass
pattern. Indeed, as discussed below, the corresponding $\d^{RR}$ term
is somewhat suppressed by negative interference and, generically, the
$\d^{LR}$ term is expected to be proportional to a lepton mass, and
furthermore it is of $\tilde{B}$ origin, less enhanced for relatively
small $|\mu |$ values than the $\d^{LL}$ one. Thus, a measurement of,
\eg , the $\meg$ decay could be interpreted as a measurement of
$\d^{LL}_{12} \, $ for this mass pattern - and not only an upper bound
- if supersymmetry is assumed, or discovered.

\subsection{Limits on $\delta^{RR}$}
 
As shown in (\ref{all}) in  the Appendix, the coefficient  of the
$\d^{RR}$'s gets only two $U(1)$ contributions, $I'_{B(R)} - I'_R\, ,$
with opposite signs - a model independent result that follows from the
sign of the hypercharge. Therefore, they can compensate each other in
some region of the parameter space, where the limits on the
$\delta^{RR}$'s thus become very weak, as we now turn to discuss.  Let us
write again the approximations in the Appendix as follows:
\beq
I'_{B,R}-I'_R \approx \frac{\mu ^2}{2\bar{m} ^4} 
\left(h_1(\bar{x})+k_1(\bar{x})\right) -\frac{1}{m_R^2} h_1(x_R) 
\label{approxdRR}
\eeq 
\noindent Then the ratio $I'_{B,R} / I'_R$ is $O(1)$ for $\mu^2 m_R^2
\approx \bar{m} ^4 \, ,$ which occurs in mSUGRA for $m_R \approx 6
M_1\, .$ The exact results are shown in fig. \ref{rappamp} b). More
generally, one always has a sector in the parameter space, roughly for
$|\mu |\approx m_R\, ,$ where the $\d^{RR}$'s remain unconstrained or
are poorly constrained.

The constraints for $\d^{RR}_{12}$  are shown in fig. \ref{4d12} where
they clearly become mediocre in the sector where $|\mu |\approx m_R\,
.$ The same appears in the bounds on $\d^{RR}_{23}$ shown in fig.
\ref{4d23}. As in the previous section, to read the present limit on
$\d^{RR}_{23}$ and $\d^{RR}_{13}$, the values on the isocurves of fig.
\ref{4d23} have to be increased by a factor of $30$ and $50\, ,$
respectively. Thus, by now $\delta^{RR}_{23}$, $\delta^{RR}_{13}$ are
not constrained at all.


\subsection{Limits on $\d^{LR}$}

Only the pure $\ti B$ graph contributes to $\delta^{LR,RL}_{12}$ and
$\delta^{LR,RL}_{23}$ .  Their upper limits are displayed in figs.
\ref{4d12} and \ref{4d23} and they feature a typical $\ti B$ shape in
the $(m_R, M_1)$ semi-plane. These limits are quite small for
$\meg$ and the proposed improvement would lower them by a factor of
30.    For the present sensitivity to $\tmg$ ($\teg$), the numbers on
the isocurves have just to be increased by a factor of $30$ ($50\, ,$
respectively):  it follows that $\delta^{LR,RL}_{23(13)}$ seems already
significantly constrained. The shape of the bounds becomes more
transparent if one adopts for the $\delta^{LR}$ term the approximation
suggested in (\ref{all}) in the Appendix:
\bea
\frac{m_R m_L}{\mu m_j \tan\beta} I_B & \approx &
\frac{\mu}{m_R m_L m_j \tan\beta}\frac{1}{x_L - x_R} 
\left( x_L\, g_1(x_L) - x_R\, g_1(x_R) \right) \nn \\
& \approx &\frac{\mu}{m_j \bar{m}^2 \tan\beta}h_1(\bar{x}) 
\ , \label{approxdLR}
\eea
\noindent where the coefficient in front of $I_B$ normalizes 
this term like all others in (\ref{all}). 

The smallness of the bounds should not be misapprehended since, as
already recalled, the $\delta^{LR}$'s must be proportional to the Higgs
v.e.v., hence to the relevant lepton masses on generic grounds, so that
they naturally are  at most $O(m_j / m_{susy} ).$ The factor
$\frac{\mu}{m_j\tan\beta}$ is taking this fact in account and
strengthening the bounds with respect to the chirality non-flip ones.
Note that the bounds are not proportional to $\tan\beta$ here.


\section{MDM and EDM} \label{sec:medm}
 
If the discrepancy between the experimental value of the muon anomalous MDM 
\cite{damuexp} 
and the SM prediction 
(see for instance \cite{SMpred} and references therein) 
turns out to be significant, it would signalize new physics, possibly supersymmetry 
\cite{amusus1, amusus2, ell}. 
Actually, the uncertainties are quite close to the level of the contributions
that are generically predicted in supersymmetric theories. On the
contrary, EDM does not suffer from this problem, since the SM
contribution turns out to be far below the potential supersymmetric
contribution.  By now, only $d_e$ gives interesting constraints, while
$d_\mu$ and $d_\tau$ are not yet able to provide significant constraints 
\cite{dsus1, dsus2}.  
However, it has been proposed to increase the sensitivity
to $d_\mu$ by six and even eight orders of magnitude.  This improvement
could provide interesting informations to be discussed later on.

In this section we briefly reappraise the constraints on the flavour and
CP dependence of the slepton masses coming from the present data on the
leptonic MDM and EDM and those prognosticated in future experiments.
The supersymmetric contributions to the $a_i\, ,d_i\  (i=e\, ,\mu\,
,\tau)\, ,$ are given by the diagonal elements of (\ref{all}) which
read:
\bea
&&\frac{a_i}{m_i} + \frac{2id_i}{e} = \frac{\alpha M_1 } 
{4\pi |\mu|^2 \cos^2\theta _W}
\left[\  m_i\mu \tan\beta(I_B + \frac{1}{2}I_L - I_R + I_2 )
-  A_i^* m_i\, I_B \right. \ \ \ \ \ \ \ \nn \\
&& +\left. m_R m_L \left( \delta^{LL}_{ik} \delta^{LR}_{ki}\, I'_{B,L}
+ \delta^{LR}_{ik} \delta^{RR}_{ki}\, I'_{B,R}\right)
+ m_k \tan\beta \left(\delta^{LL}_{ik}\delta^{RR}_{ki}\eta^*_k \mu + 
\delta^{LR}_{ik}\delta^{LR}_{ki} \eta_k \mu^* \right)
 I''_B \frac{}{}\right] \ \ \
\label{mdmedm}
\eea 
\noindent where we identify two kinds of terms:
\begin{itemize} 
\item[ i)] FC contributions arising only from the slepton mass
parameters that are flavour conserving (in the lepton basis)
\item[ ii)] FV contributions involving flavour non-diagonal parameters
in the mass matrices, \ie , misalignment.
\end{itemize}

The first line in (\ref{mdmedm}) contains the terms of FC type, while
the FV ones appear in the second line. The latter depend on products of
$\delta$'s. All the terms have to be proportional to a lepton mass as
emphasized before. However, by their own nature, the FC terms are
proportional to the mass of the same fermion, leading to the
proportionality  between EDM and lepton masses, $d_i \propto m_i\, ,$
and between MDM and the squared lepton masses, $a_i \propto m_i^2\, .$
In the limit where all the slepton masses are flavour independent one
has the FC relations:
\beq
\frac{d_i}{d_j}=\frac{m_i}{m_j}\, , \qquad \qquad 
\frac{a_i}{a_j}=\frac{m_i^2}{m_j^2} \, .\label{naive}
\eeq
\noindent In the development that leads to (\ref{mdmedm}), this is
satisfied by the first term and the second FC contribution only
violates it by a factor ${A_i}/{A_j}\, .$ This violation is expected
to be too small to significantly change the hierarchy driven by the
lepton masses in (\ref{naive}), with the important exception of the
LFV and CP phases due to the RGE running
\cite{ellis, imsusy02, ms2}. 
If these were the main
contribution to EDM, $d_\mu$ should not exceed $3 \cdot 10^{-25}$ e cm,
due to the present bound on $d_e$.  Notice that this value roughly
corresponds to the planned sensitivity for $d_\mu$.

\begin{figure}[!h]
\centerline{\psfig{file=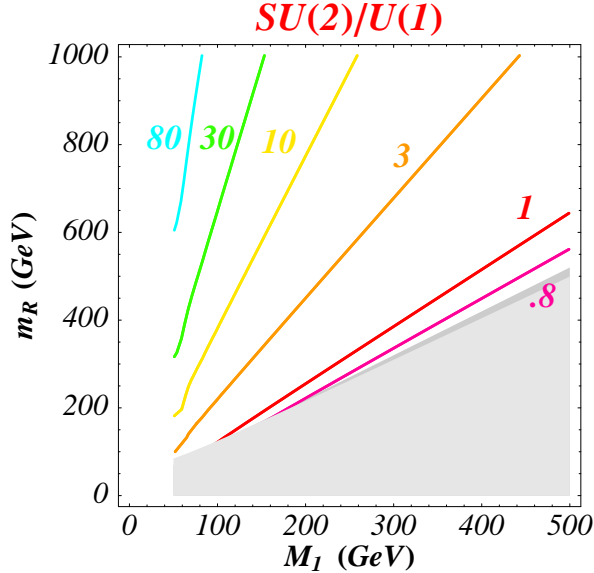,width=0.5\textwidth}}
\caption{Ratio of the $SU(2)$ and $U(1)$ contributions in mSUGRA. 
Notice that it is independent on $\tan\beta$.}
\label{PL_amusu2u1}
\end{figure}

Instead, some of the FV terms are proportional to a different, possibly
heavier lepton mass that break the relation (\ref{naive}) and the
corresponding hierarchy in the leptonic MDM/EDM. Thus, even if the FV
terms in (\ref{mdmedm}) possess two factors $\d $'s, they are boosted
by $m_k/m_i$ and could bypass the mass scaling (\ref{naive}), as already discussed in 
\cite{fms, romstr}
\footnote{And, before, for the quark sector, in 
\cite{bs}
.}. Therefore, the observation of $d_\mu$ above $3
\cdot 10^{-25}$ e cm is still a realistic possibility that deserves
experimental tests, also because the breaking of the mass scaling
rule would be a basic fact in supersymmetric CP violations.

In any case, the experimental limits on the leptonic MDM/EDM shall put
upper bounds on the various FC and FV parameters - barring weird
cancellations among them. Conversely, the limits on the LFV transitions
put limits on the parameters in the FV part of MDM/EDM. We now turn to
discuss these limits and their interplay.


\subsection{FC contribution to MDM and EDM }\label{subsec:fcmedm}

\begin{figure}[!h]
\centerline{\psfig{file=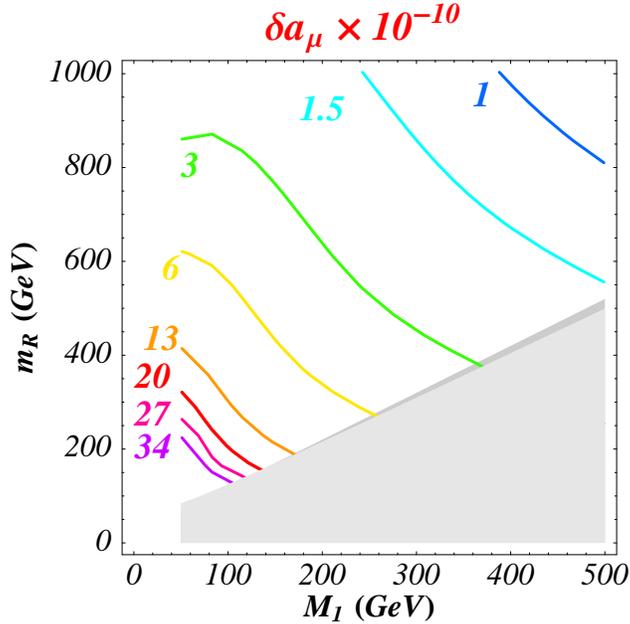,width=0.5\textwidth}}
\caption{Value of $\delta a_\mu$ for $\tan\beta=10$ in mSUGRA.}
\label{PL_amu}
\end{figure}

The FC contributions have been discussed in many recent papers for both
the leptonic MDM \cite{amusus1, amusus2} and EDM \cite{dsus1, dsus2}. It has
been realized that these contributions can be suppressed only by
increasing the supersymmetry breaking scale, with the exception of the
following two contrived situations far away from the usual mSUGRA
constraints: (i) $M_2/M_1 < 0$, meaning that $SU(2)$ and $U(1)$ are not
simply unified, a possibility offered by some brane models \cite{kane};
(ii) a cancellation between the terms with $\mu\tgb$ and $A\, ,$
involving parameters of different nature. Up to these somewhat
contrived possibilities, one can constrain each term separately, as we
now discuss.

The main characteristics of the results can be understood from fig.
\ref{PL_amusu2u1} where the $SU(2)$ and the $U(1)$ contributions are
compared. They have the same sign if $M_2/M_1 > 0$. The $\ti B$
graph becomes comparable to the chargino one only in the region where
$\mu^2 \gg m_R^2\, ,$ \ie , $M_1^2 \sim O(m_R^2)$ in mSUGRA.  In this
model, the $SU(2)$ and the $U(1)$ contributions become equal for  $M_1
\approx .72 m_R\, .$ The chargino dominates elsewhere.  The
approximations given in the Appendix perfectly describe this pattern.
Note that only the pure $\ti B$ graph is relevant for the $A-$term
input.

\begin{figure}[!t]
\centerline{\psfig{file=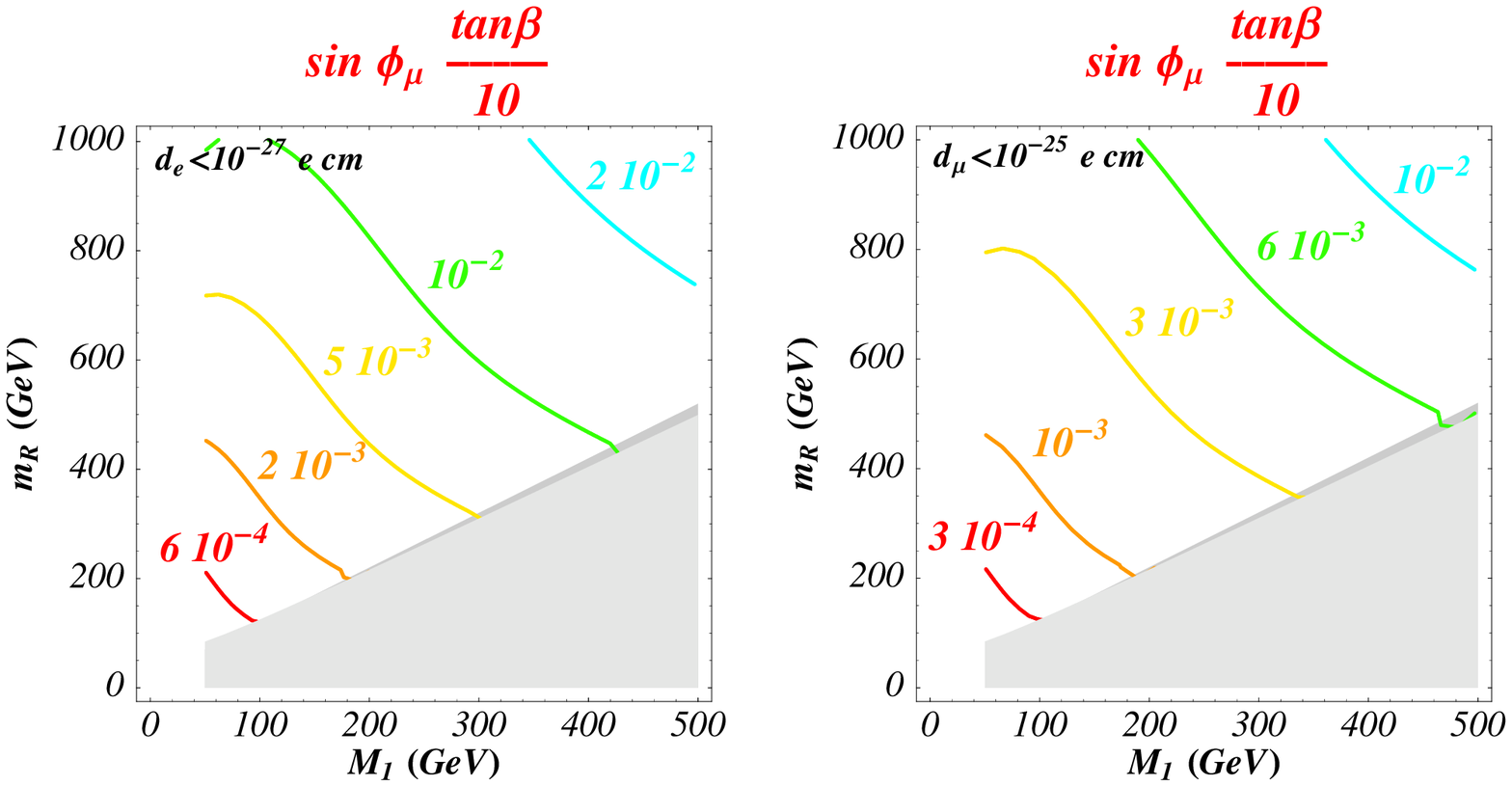,width=0.9\textwidth}}
\caption{Upper limit on $\sin \phi_\mu \tan\beta/10$ in mSUGRA. Vanishing $A$-term is assumed.}
\label{PL_phmu}
\vskip 2 cm
\centerline{\psfig{file=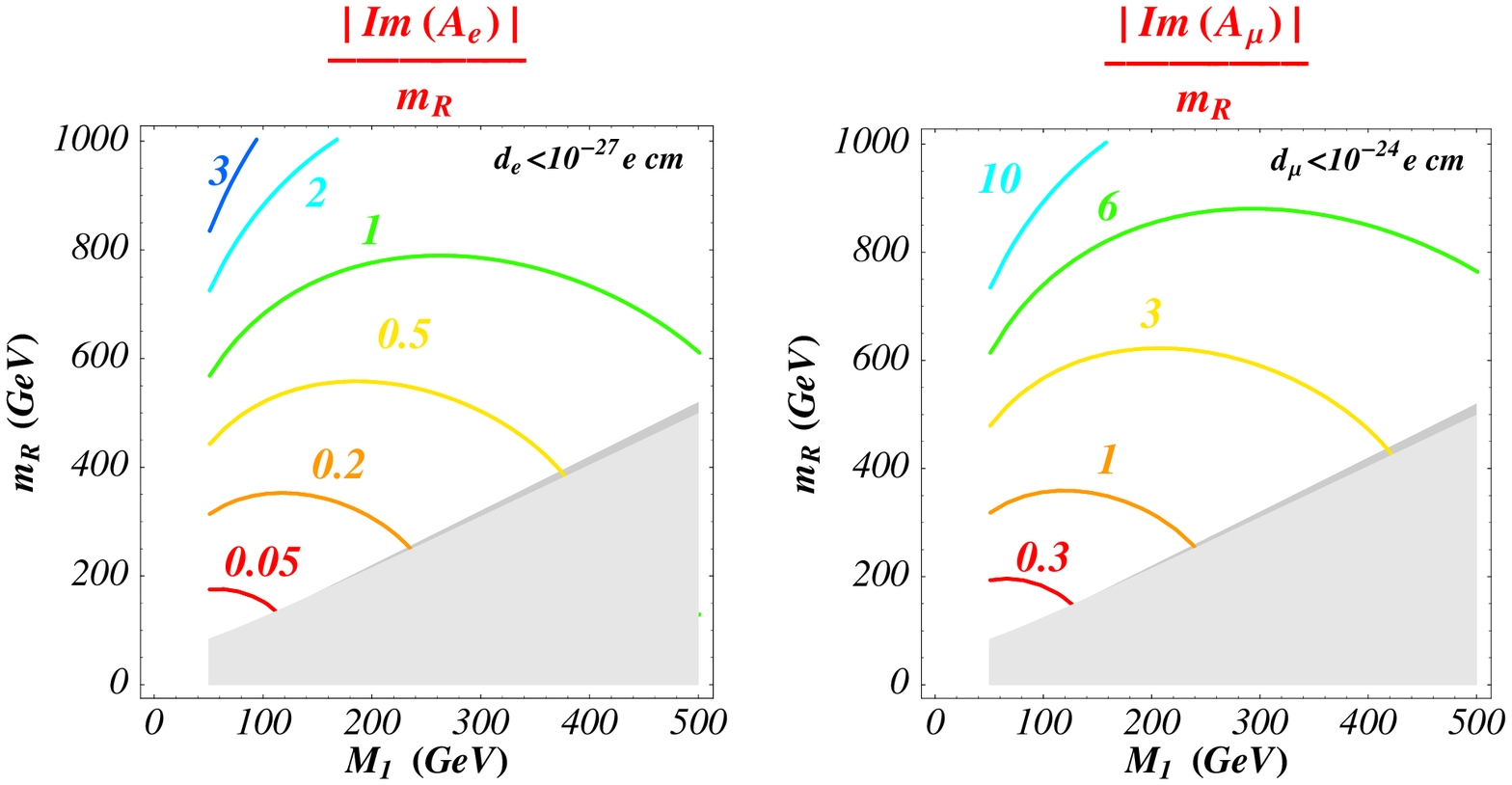,width=0.9\textwidth}}
\caption{Upper limit on $|{\rm Im} A_e| / m_R$ and  $|{\rm Im} A_{\mu}|/ m_R$.}
\vskip 1 cm
\label{PL_phA}
\end{figure}

We refer to the abundant literature for the MDM but for completeness we
plot in fig. \ref{PL_amu} the contour lines for $a_{\mu} \, .$ Since
the naive mass scaling is naturally realized for the FC parameters,
$a_{e} = m_e^2 a_{\mu}/ m_{\mu}^2$ and $a_{\tau} = m_{\tau}^2 a_{\mu}/m_{\mu}^2
\, .$ The $A-$term is marginal in this result which corresponds to
$\mu\tgb \gg A\, .$ When the present theoretical and experimental
uncertainties will settle down, this plot will provide interesting 
constraints. Nevertheless, it should be noted that in the pure chargino
approximation one would overlook the constraints close to the dark matter
preferred region where the $\ti B$ is important.

As for the leptonic EDM, the upper bounds are separately shown
for the two FC terms in (\ref{mdmedm}). We choose phases in such a way
that $M_1$ is real (and so is $M_2$ unless we state the contrary).
The relevant phases are then $\phi_{\mu}$ and $\phi_A\, ,$ respectively.
The limits on $\phi_{\mu}$ are shown in fig. \ref{PL_phmu} for the present
limits on $d_e$ and they clearly illustrate the so-called supersymmetric
CP problem, the experimental requirement of very small CP phases in the
soft mass parameters as compared, \eg , with the Kobaiashi-Maskawa one.
The other plot in fig. \ref{PL_phmu} shows that even with a considerable
upgrading of the limits on $d_{\mu}$ one cannot significantly improve
the present results on $\phi_{\mu}\, .$ Instead the much better precision
in future searches for $d_e$ would bring these limits down to extremely
small figures.

Let us now assume that $\phi_{\mu}$ satisfies these requirements, and
look for the limits on ${\rm Im} A$ that would be obtained with the degree
of precision of existing data for $d_e$ and the one that  has been 
advertised for projected experiments on $d_{\mu}$. This is
shown in fig. \ref{PL_phA} for ${\rm Im} A_e / m_R$ and 
for ${\rm Im} A_{\mu}/ m_R$. The curves exhibit a typical $\ti B -$like shape,
since this is the only contribution. There are well-known model
dependent upper bounds on these parameters to avoid colour and e.m.
charge breaking, roughly $|A_i |/ m_R < 3$ in mSUGRA. The limits shown
in these figures are already much better.

\subsection{FV contributions to EDM and MDM}\label{subsec:fvmedm}

\begin{figure}[!b]
\centerline{\psfig{file=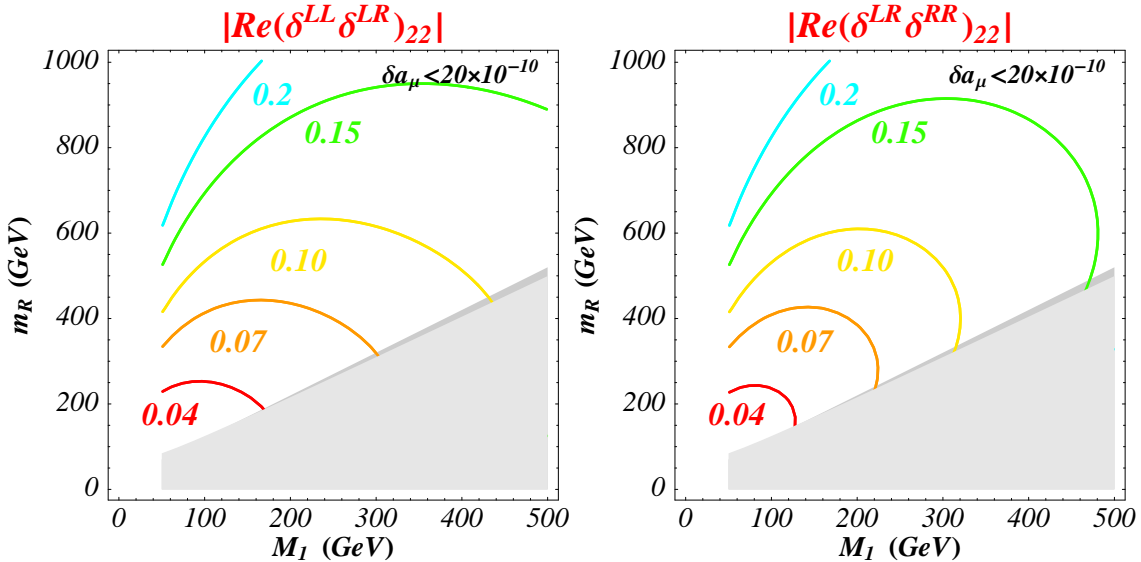,width=0.9\textwidth}}
\caption{Upper bounds on the real part of products 
of $\d$'s obtained by assuming a limit of $20\times 10^{-10}$ on the
supersymmetric contribution to $a_{\mu}$.}
\label{PL_Re2d}
\end{figure}

There are four sums of products of $\d $'s that are constrained
by the limits on MDM and EDM : 
\beq
\delta^{LL}_{ik}\delta^{RR}_{ki} \qquad
\delta^{LR}_{ik} \delta^{RR}_{ki}\qquad \delta^{LL}_{ik} \delta^{LR}_{ki}
\qquad \delta^{LR}_{ik}\delta^{LR}_{ki} \ .\label{3dd}
\eeq 
They are all obtained from the multi-insertion development of the
slepton propagator in the pure $\ti B$ graph. Therefore their
coefficients are roughly of the same order of magnitude. Approximations
are offered in the Appendix that allow for a quick adaptation to other
mass configurations. Note that their coefficients include a lepton mass
but for $\delta^{LR}_{ik} \delta^{RR}_{ki}$ and $\delta^{LL}_{ik} \delta^{LR}_{ki}$.  
However, as already stressed, such a factor should be enclosed in $\delta^{LR}_{ik}\, .$ Each
product would have two terms since, by assumption, $i \neq k\, .$
However, barring any fortuitous cancellation, each term is constrained
by the experimental bounds. We concentrate on those with $k =\tau$ due
to the $m_{\tau}$ factor and because the associated $\d $'s are less
constrained by LFV decays. Hence we define $\phi _{\tau}$ as the phase
of $(\mu^*\tgb - A_{\tau}) ~,$ close to the phase of $\mu $ for large
$\tgb ~.$

\begin{figure}[!t]
\centerline{\psfig{file=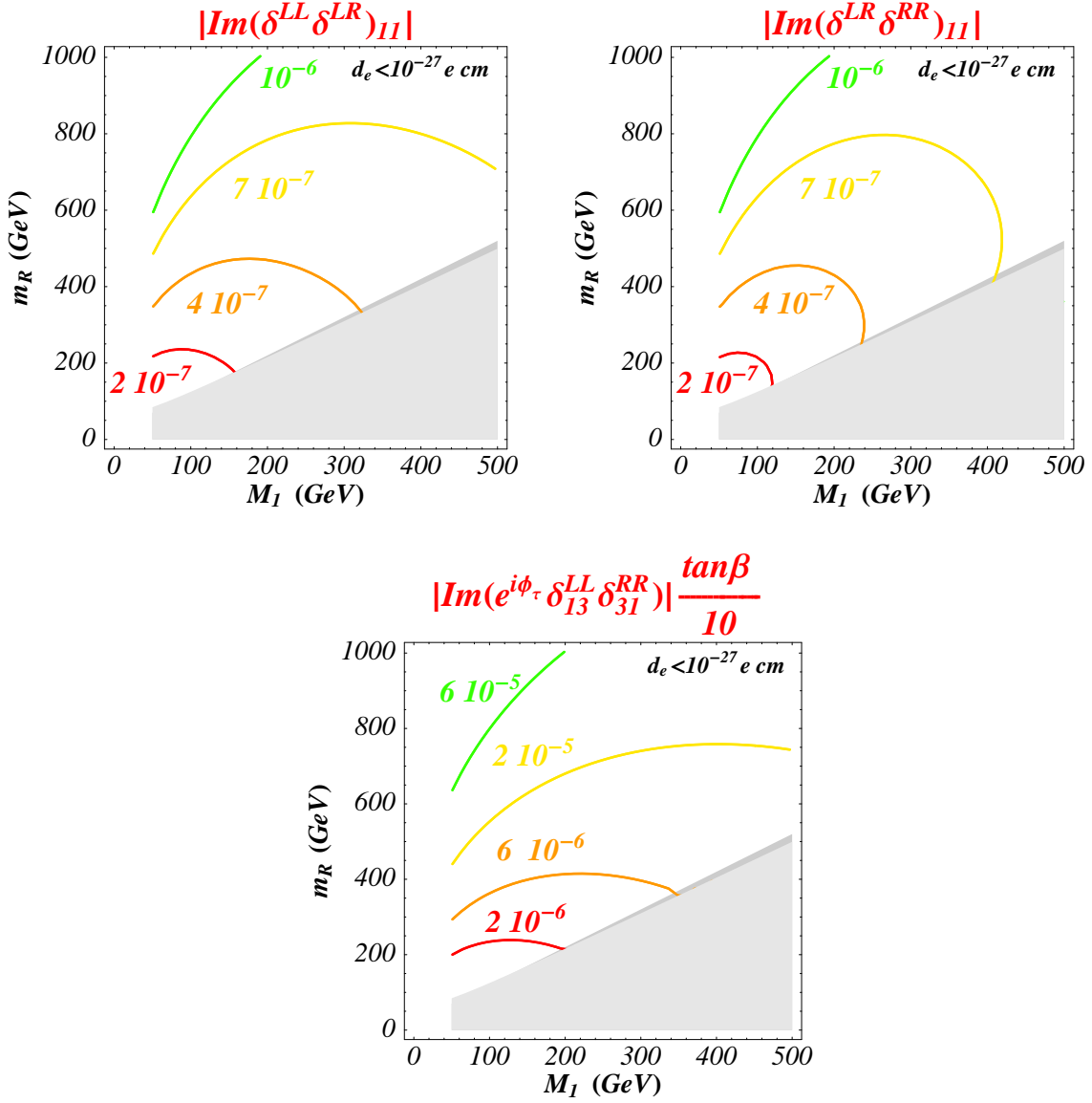,width=0.9\textwidth}}
\caption{Upper bounds on the imaginary part of products of 
$\d$'s with the present limit on $d_e$; $\phi_{\tau}$
is defined in the text.}
\label{PL_2de}
\end{figure}

\begin{figure}[!ht]
\centerline{\psfig{file=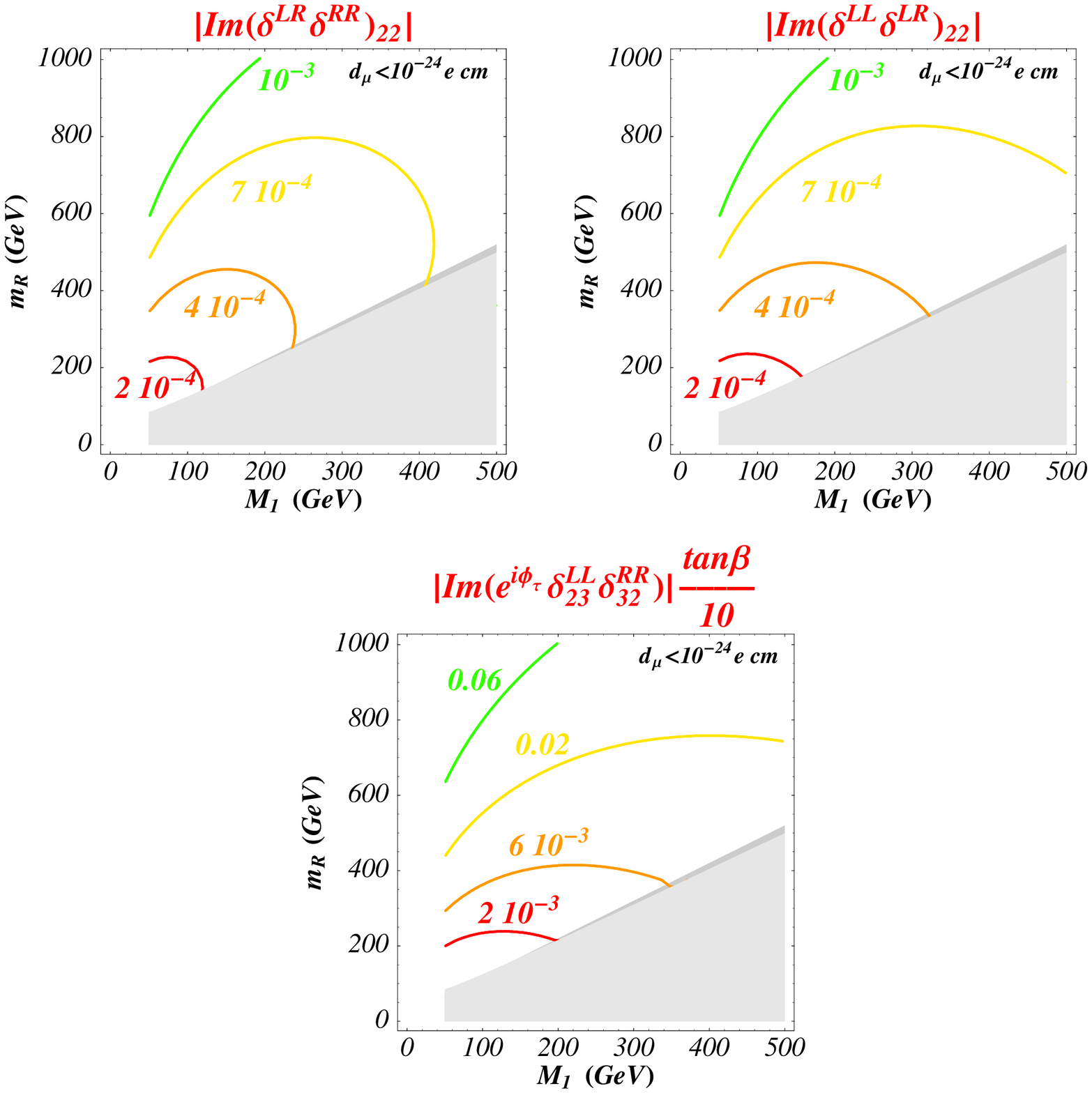,width=0.9\textwidth}}
\caption{Upper bounds on the imaginary part of products of 
$\d$'s for a limit of $10^{-24}$e\, cm on $d_{\mu}$; $\phi_{\tau}$
is defined in the text.}
\label{PL_2dmu}
\end{figure}

The bounds obtained on the real parts by taking into account the
present uncertainties in $a_{\mu } $ are shown in fig. \ref{PL_Re2d}.
Presumably, one cannot do much better because of the level of
theoretical uncertainties. Yet, as compared with the results in fig.
\ref{4d23}, rescaled to the present experimental bounds on $\tmg$ and
$\teg\, ,$ these products are useful. In particular, they are not
proportional to $\mu\tgb\, .$

The bounds on the imaginary parts derived from the experimental limits
on EDM are free of SM contributions, and depend only on the experimental accuracy. 
For $d_e$ we consider the existent bound and derive the curves in fig. \ref{PL_2de} 
which put limits on the imaginary part of $\delta^{LL}_{13} e^{i \phi_\tau}\delta^{RR}_{31}\, ,
\delta^{LR}_{13} \delta^{RR}_{31}$ and $\delta^{LL}_{13}\delta^{LR}_{31}\, .$ 
For $d_{\mu } $ we consider the projected precision of $10^{-24}$ e cm and show the results in fig. 
\ref{PL_2dmu}. 
The limits on $Im (e^{-i\phi _{\tau}} \d ^{LR}_{i3} \d ^{LR}_{3i})$ are not displayed because
they are the same as those on $Im (e^{i\phi _{\tau}} \d ^{LL}_{i3} \d ^{RR}_{3i})$. 
The limits are relatively stringent, even when those involving a $\d ^{LR}$ are increased 
by a factor $m_R/m_{\tau}$ to extract the lepton mass dependence. 
Notice that the limits on $Im (e^{i\phi _{\tau}} \d ^{LL}_{i3} \d ^{RR}_{3i})$ 
and $Im (e^{-i\phi _{\tau}} \d ^{LR}_{i3} \d ^{LR}_{3i})$ are inversely proportional
to $\mu\tgb.$ 

All these results are easily understood from the approximations
in the Appendix, which are also useful for a quick evaluation of
alternative models. 


\section{Limits on FV contributions to EDM from LFV decays}
\label{subsec:edmmax}

Of course, all these tests of the lepton flavour structure of the soft
parameters of supersymmetric extensions of the SM are quite
complementary. For instance, as we now turn to discuss, the conjunction
of experimental bounds on LFV transitions and on MDM and EDM would help
in disentangling the FC and FV contributions in (\ref{mdmedm}) and in
learning whether CP violation is more present in one or the other kind
of soft masses.

As a case study, we concentrate here on $d_{\mu}$ and evaluate the
maximal FV contribution by using the limits on $|\d ^{LL}|\, , |\d^ {RR}|
\, , |\d ^{LR}|\, ,$ matrix elements obtained from $\tmg $ (those
from $\meg $ are much smaller and can be neglected). 
Let us rewrite the sum of the moduli of
FV terms contributing to $d_{\mu}$, keeping only $k =\tau\, :$ 
\bea
\frac{2d_{\mu}}{e} &\leq & \frac{\alpha M_1 } {4\pi|\mu |^2\cos^2\theta _W}
\left[\ m_R m_L\left( |\delta^{LL}_{23}\delta^{LR}_{32}|\, I'_{B,L}
+ |\delta^{LR}_{23} \delta^{RR}_{32}|\, I'_{B,R}\right) \right. \nn \\
&+ &\left. \left( |\delta^{LL}_{23}\delta^{RR}_{32}| + 
|\delta^{LR}_{23} \delta^{LR}_{32}| \right)
|\mu\tan\beta - A^*_{\tau}| m_{\tau} I''_B  \frac{}{} \right]
\label{maxi}
\eea 
This limit is conservative since we replace
the imaginary part of the sum by the sum of the moduli.
As shown in the Appendix, the integrals $I'_{B,L}~ , ~ I'_{B,R}~ , ~
I''_B~ ,$ are all of the same order of magnitude. In particular, for $m_R > M_1$,
it can even be approximated in mSUGRA or similar models as
\bea
\frac{2d_{\mu}}{e} &\leq & 
\frac{\alpha M_1 }{8\pi\bar{m}^2\cos^2\theta _W}
\left[\frac{}{}\ \left( |\delta^{LL}_{23}\delta^{LR}_{32}|
+ |\delta^{LR}_{23} \delta^{RR}_{32}|\right)  \right. \nn \\
& & \left. + 2\left( |\delta^{LL}_{23}\delta^{RR}_{32}| + 
|\delta^{LR}_{23} \delta^{LR}_{32}| \right)
|\mu\tan\beta - A^*_{\tau}| \frac{m_{\tau}}{\bar{m}^2} \right]\, 
h_1(\bar{x})
\label{approximax}
\eea 
where $\bar m^2= (m_L^2 + m_R^2)/2$ and $\bar x = M_1^2/\bar m^2$.

\begin{figure}[!b]
\centerline{\psfig{file=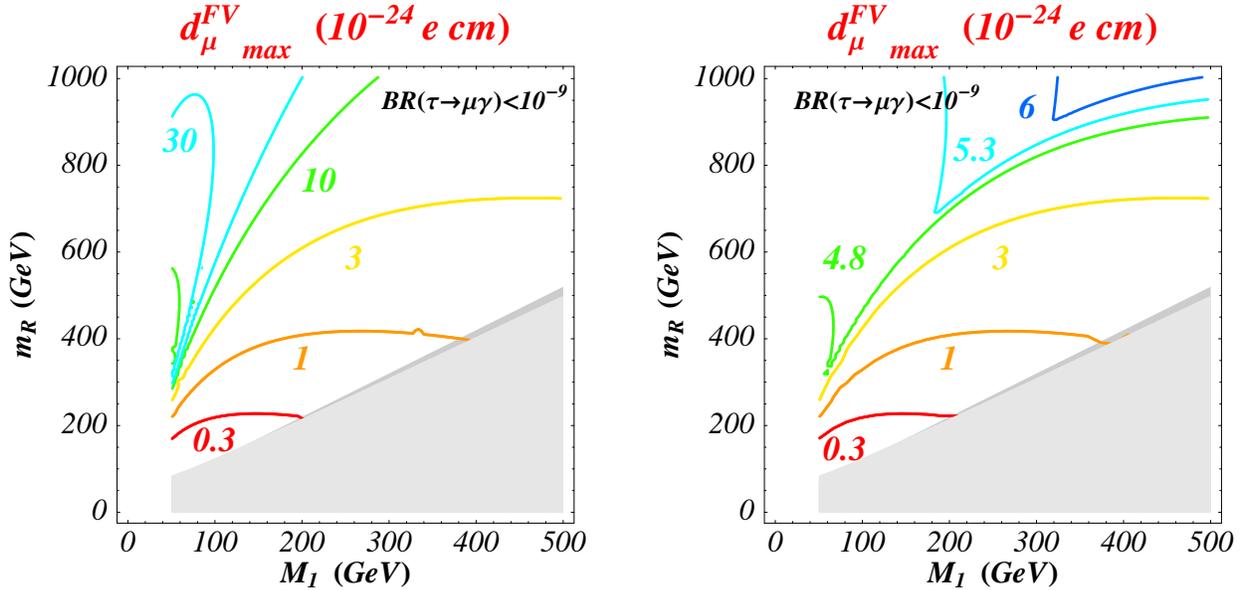,width=1\textwidth}}
\caption{Upper bounds on the FV contribution to $d_\mu$ 
assuming a limit on $\tmg$ at the level of $10^{-9}$. In the 
left figure the limits on $\d ^{RR}_{23}$ are those from fig. \ref{4d23}
while in the right figure we put an additional upper limit of $0.5$
on them to take into account the way they are defined. Analogous results 
for $d_e$ are discussed in the text.}
\label{PL_dmuFVmax}
\end{figure}

In section 3, limits on the $|\d_{23}|$'s were obtained from the experimental bound 
on ${\rm BR}(\tmg)$  barring important cancellations among the different contributions,
\ie~ from eq. (\ref{lfvbis}). 
In the same spirit, we consider the maximum of the r.h.s.
of eq. (\ref{maxi}) with eq. (\ref{lfvbis}) as a constraint. 
Actually, since $|\delta^{RR}_{32}|$ is less
constrained than the others, the bound on $d_{\mu}$ come mostly from
the term with the product $|\delta^{LR}_{23}\delta^{RR}_{32}|~ .$ 
In fig. \ref{PL_dmuFVmax} we have illustrated the kind of bounds on
$d_{\mu}$ that are obtained as described, by assuming an experimental
limit $BR(\tmg ) < 10^{-9}~ .$ 
As appears from fig. \ref{4d23}, even with such a sensitivity the limits on $|\delta^{RR}_{32}|$ 
are meaningless in the sector of the $(m_R, M_1)$ plane where they are larger than one.
The left plot has been obtained by substituting the limits on $|\delta^{RR}_{32}|$ 
of fig. \ref{4d23}, while the plot on the right is corrected for by the additional condition
$|\delta^{RR}_{32}| < 1/2 ~$. 
The latter plot is actually relevant for $BR(\tmg)$ at the level $10^{-9}$,
while the first plot is suitable for a quick adaptation of the limits to other levels
of sensitivity.
Notice that a sensitivity to $\tmg$ at the level of $10^{-9}$ would push down $d_\mu^{FV}$, 
close to the $d_{\mu}$ values that could be tested by the planned experiments 
in the near future.  

If we take the present experimental limit $BR(\tmg ) < 10^{-6}~ ,$ 
together with a limit of $1/2$ on all the $\d $'s, we basically get
an upper bound on $d_\mu$ of the order of few $10^{-22}$ e cm.
Therefore we conclude that at present there is still plenty of place for a LFV
contribution to $d_\mu$ up to three orders of magnitude larger than
$m_\mu/m_e d_e$. 
 
For $d_e$ there are two FV contributions to be bounded, corresponding
to intermediate $\ti \mu$ and $\ti \tau$, respectively. The the limits
on the former can be read from those in  fig. \ref{PL_dmuFVmax} by  
multiplying them by $ 0.2\times 10^9 ~{\rm BR}(\meg)$ and are close to
the present experimental bounds on $d_e$ for the existent limits on the
$\meg$  decay.  Instead, for the latter the limits come out much worse,
at the level of $10^9 ~{\rm BR}(\teg)$, namely of those on $d_{\mu}$,
since the bounds on $\teg$ and $\tmg$ are now very close. 



\section{Concluding Remarks}\label{sec:conclusions}

The whole set of experimental constraints on the flavour structure of
the sleptons masses (in the basis where all interactions are flavour
diagonal) are gathered here and displayed in such a way to allow for a
ready understanding at both a qualitative and a quantitative level.
The bounds on $\meg $ transitions are already very constraining in the
$\mu - e$ sector and the prospects for the near future are encouraging.
As for $\tmg$, the present data are restrictive only at the low side of
the sparticle mass spectrum so that any improvement would be
particularly welcome.

The searches for the electric dipole moments of the electron and of the
muon provide a unique information on the CP violation in the lepton
sector. The present data already point to a supersymmetric CP  problem
similar to what is encountered in the squark sector. The simultaneous
analysis of LFV transitions and of the lepton EDM would help
discriminating between a CP violation in the flavour blind portion of
the supersymmetry breaking parameters or in the presumably richer
flavour dependent one. Again, this would require a solid upgrading in
the $\tmg $ and $d_\mu $ searches to ameliorate the present bounds by a
few orders of magnitude.

The level of phenomenological upper bounds on the misalignment
parameters, $\d $'s, that should be reachable in a near future would
provide an indirect test of the existence of two kinds of fundamental
particles that are too heavy to be more manifest:  the right-handed
neutrinos from the seesaw mechanism and the triplet partners of the
Higgses partners from GUTs.  In supersymmetric theories the radiative
corrections due to these particles before their decoupling could leave
their footprints in the flavour structure of the supersymmetry breaking
masses. They are suppressed by loop factors and by the strength of the
Yukawa couplings involved, but only logarithmically in their masses.
The level of precision of the LFV and even the lepton  EDM searches
should draw near the one needed to test Yukawa couplings and mass
scales of the supersymmetric seesaw and GUT heavy states \cite{ellis,
ell, lms1, lfvss, ms2, barb}.

At the present and near future level of the experimental precision, the
observation of any of the LFV and CP violating transitions discussed in
this paper would point to a characteristic scale for these processes
around that conjectured for low energy supersymmetry. This is very
different from the measured FCNC and CP violating quark transitions
that are primarily SM processes, and from the large seesaw scale
suggested in the observed LFV in neutrino  oscillations. But even a
substantial improvement in the upper bounds on any of these processes,
especially  $\tmg $ decay, would provide decisive data about the
flavour and CP violation dependences of the supersymmetry breaking
parameters and their origin.

\vskip 1cm
\noindent {\it Acknowledgement}: We thank O.Vives for pointing out a missing numerical 
factor in a previous version of the lower plot of fig.2 and some misprint in appendix A.


\appendix

\section{Appendix}

\subsection {FC and FV Amplitudes}

In this Appendix we  exhibit the (multiple) insertion approximation
for the FC and the FV dipole moment transition amplitude, as discussed
in Section 2, where the approximations are defined and justified.
They are displayed so as to make more explicit the main
dependence on the soft mass parameters and to facilitate a qualitative
understanding of the numerical results. The contributions from the
different diagrams displayed in Section 2 are separated. They are indicated by the
lower indices: $B$ for the pure $\tilde{B}$ diagram, $L$ and $R$ for the
$\tilde{B}\, -\, \tilde{H}^0$ one with $L$ and $R$ sleptons, respectively,
and $2$ for the $\tilde{W}\, -\, \tilde{H}$ ($SU(2)$) one. 
\bea
M_{ij}  =  2(\mu_{ij} + i\frac{d_{ij}}{e}) 
&=&
 \frac{\alpha M_1 }{4 \pi |\mu |^2 \cos^2 \theta _W} 
\ \left\{ \  \left[  
\delta_{ij} m_j(\eta_j I_B + \frac{1}{2} I_L - I_R + I_2 ) 
\right. \right.\nn \\  
& & -\ \delta^{LL}_{ij}m_i (\eta_i I'_{B,L} + \frac{1}{2} I'_L  + I'_2 ) 
\ - \ \delta^{RR}_{ij} m_j ( \eta_j I'_{B,R} - I'_R )  \nn \\
& & +\ \left. ( \delta^{RR}_{ik} \eta_k \delta^{LL}_{kj} 
+ \delta^{RL}_{ik}\eta^*_k \frac {\mu}{\mu^*}\delta^{RL}_{kj} ) m_k I''_B 
\right]\mu^* \tgb \nn \\
& & -\ \biggl.  
\Bigl[ \delta^{RL}_{ij}I_B - \delta^{RR}_{ik}\delta^{RL}_{kj} I'_{B,R}  
- \delta^{RL}_{ik}\delta^{LL}_{kj} I'_{B,L} \Bigr]\, m_R m_L\ 
+ \ \dots \biggr\}  \label{all} 
\eea
\noindent where,
\beq
\eta _j = \left( 1 -\frac{A_j}{\mu^* \tan \beta} \right)
\label{Mapp}
\eeq
\noindent and terms that are less relevant or higher order are
omitted, while we keep terms that could break the proportionality between the
dipole moments and the lepton masses for the electron and the muon.
In order to define the functions I (\ref{all}), it is convenient to introduce
the new variables:
\beq
x_L= \frac{M_1^2}{m_L^2}
\qquad x_R= \frac{M_1^2}{m_R^2} \qquad x'_L= \frac{|M_2|^2}{m_L^2} 
\qquad  y_L= \frac{|\mu^2|}{m_L^2} \qquad y_R= \frac{|\mu^2|}{m_R^2}
\eeq
The dependence of the reduced amplitudes $I$'s on the mass parameters is as 
follows:

\bea
I_L (m_{L}^2,\, M_1^2,\, \mu^2 ) & = & \frac{1}{m_{L}^2}
\frac{y_L}{y_L - x_L} \left[\, g_1 \left( x_L\right) 
- \, g_1 \left( y_L\right)\right] \nn \\ 
I_R (m_{R}^2,\, M_1^2,\, \mu^2 ) & = & \frac{1}{m_{R}^2}
\frac{y_R}{y_R - x_R} \left[\, g_1 \left( x_R\right) 
- \, g_1 \left( y_R\right)\right] \nn \\
I_2 (m_{L}^2,\, M_2^2,\, \mu^2 ) & =  &
\frac{M_2\cot ^2\theta_W}{M_1 m_{L}^2} \frac{y_L}{y_L - x'_L} 
\left[\, g_2 \left( x'_L\right) 
- \, g_2 \left( y_L\right)\right] \nn \\ 
I_B (M_1^2,\, m_{L}^2,\, m_{R}^2) & =  & 
\frac{1}{m_{R}^2 - m_{L}^2} \left[ y_L \, g_1 \left( x_L \right) 
- y_R \, g_1 \left( x_R \right) \right] 
\nn \\
I'_L (m_L^2,\, M_1^2,\, \mu^2 ) & = & \frac{1}{m_L^2}
\frac{y_L}{y_L - x_L}\left[\, h_1 \left( x_L\right)
- \, h_1 \left( y_L\right)\right] \nn \\ 
I'_R (m_R^2,\, M_1^2,\, \mu^2 ) & = & \frac{1}{m_R^2}
\frac{y_R}{y_R - x_R}\left[\, h_1 \left( x_R\right) 
- \, h_1 \left( y_R\right)\right] \nn \\ 
I'_2 (m_L^2,\, M_2^2,\, \mu^2 ) & = & \frac{M_2\cot ^2\theta_W}{M_1 m_L^2}
\frac{y_L}{y_L - x'_L} \left[\, h_2 \left( x'_L\right) 
- \, h_2 \left( y_L\right)\right] \nn \\ 
I'_{B,R} (M_1^2,\, m_L^2,\, m_{R}^2) & = & -\frac{1}{m_R^2 - m_{L}^2}
\left( y_R \, h_1\left( x_R\right)  
- m_R^2 I_B \right) \nn \\
I'_{B,L} (M_1^2,\, m_{L}^2,\, m_{R}^2) & = & -\frac{1}{m_L^2 - m_R^2} 
\left(  y_L \, h_1\left( x_L\right)  
- m_L^2 I_B \right) \nn \\
I''_B (M_1^2,\, m_{L}^2,\, m_{R}^2) & = & -\frac{m_L^2m_{R}^2}{m_L^2 - m_R^2}
\left( \frac{1}{m^2_L} I'_{B,L} - \frac{1}{m^2_R} I'_{B,R} \right)
\label{Iapp}
\eea

\noindent where the (factorized) loop integrals have the following 
expressions:
\beq \begin{array}{cc}
 g_1 (x) = \frac{1 - x^2 + 2x \ln (x) }{(1 - x)^3}   &
 g_2 (x) =  \frac{x^2-8x+7 + 2(2+x) \ln (x)}{2(x-1)^3}  \\
 h_1 (x) = \frac{1+4x-5x^2 + (2x^2 + 4x)\ln (x)}{(1 - x)^4}   &
 h_2 (x) = \frac{7x^2+4x-11 - 2(x^2 +6x+2) \ln (x) }{2(x - 1)^4} 
\end{array} \eeq

\begin{figure}[!ht]
\centerline{\psfig{file=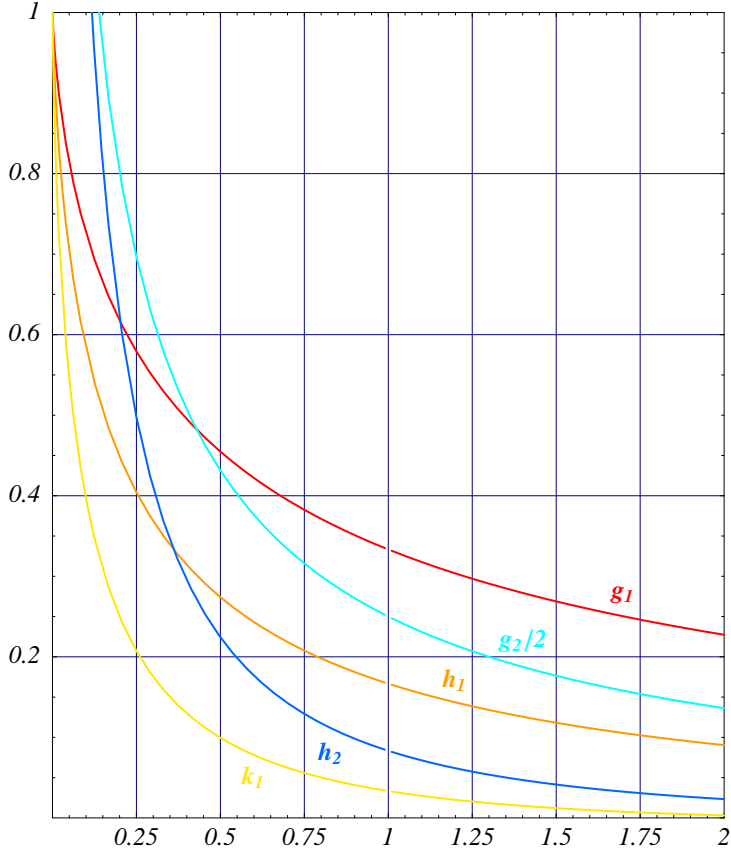,width=0.6\textwidth}}
\caption{The functions $g_i(x) ~, h_i(x)$ and $k_1(x)$.}
\label{Ffunc_ghk}
\end{figure}

These functions are plotted in Fig. \ref{Ffunc_ghk}. Note that for very
small $x,$ $\, g_1,\, \, h_1\, \rightarrow 1\, ,$ and that $g_1(1/x) =
xg_1(x)~.$ Because of the chargino mass singularity, $\, g_2,\, \,
h_2\, \sim \ln x^{-2}\,$ near the origin; this increases the relative
$SU(2)$ contributions for $ m_L^2\, \gg |M_2|^2 \, .$ Within the $SU(2)$
diagrams, only the chargino part has the logarithmic divergence and it
dominates its opposite sign $SU(2)$ neutralino counterpart everywhere.
Also, $ \, h_1 ,\, \, h_2  \, ,$ which appear in the FV amplitudes,
decrease much faster than $\, g_1,\, \, g_2$ which appear in the FC
amplitudes, since they have an additional slepton propagator in the
insertion approximation.

\subsection {Approximate Expressions}

There are several approximations that could be useful to understand
the behaviour of the FC and FV processes as shown in the figures.
In many cases they can also help to extrapolate the results in these 
figures to values of the parameters that deviate form the mSUGRA 
constraints.

First consider the case with $\mu^2\, \gg M_2^2,\, M_1^2$ which 
appears in mSUGRA and all models where $\mu^2$ is tuned to the 
gluino masses by the vacuum condition. Then, one can use the 
simplified expressions:
\bea
& & I_L \approx \frac{g_1\left( x_L\right)}{m_L^2}\, , \qquad 
I_R \approx \frac{g_1 \left( x_R\right)}{m_R^2}\, , \qquad
I_2 \approx \left( \frac{M_2\cot^2\theta_W}{M_1} \right)
\frac{g_2 \left( x'_L\right)}{m_L^2}\, , \nn \\
& & I'_L \approx \frac{h_1\left( x_L\right)}{m_L^2}\, , \qquad 
I'_R \approx \frac{h_1\left( x_R\right)}{m_R^2}\, , \qquad
I'_2 \approx  \left( \frac{M_2\cot^2\theta_W}{M_1} \right)
\frac{h_2 \left( x'_L\right)}{m_L^2}\, ,
\eea \label{approxI}
\noindent and the approximate relations:
\bea
& & I_B \approx \frac{\mu^2}{m_R^2 - m_L^2} \left( I_L - I_R \right) \qquad
I''_B \approx \frac{m_R^2 m_L^2}{ (m_R^2 - m_L^2 )^2} \left( 
y_R I'_R + y_L I'_L - 2I_B \right) \nn \\
& & I'_{B,R} \approx -\frac{m_R^2}{m_R^2 - m_L^2}\left(y_R I'_R - I_B\right)
\qquad I'_{B,L} \approx -\frac{m_L^2}{m_L^2 - m_R^2}
\left(y_L I'_L - I_B\right) \, . \label{approxI_B}
\eea

If $m_R^2$ and $m_L^2$ are not very different these expressions can 
be further approximated as,
\bea
I_B \approx\frac{\bar{y}~ h_1(\bar{x})}{\bar{m}^2}  \qquad 
I'_{B,R} \approx I'_{B,L} \approx  
\frac{\bar{y}~ (h_1(\bar{x}) + k_1(\bar{x}))}{2\bar{m} ^2}  
 \qquad I''_B \approx \frac{\bar{y}~ (h_1(\bar{x}) +2 k_1(\bar{x}))}{3\bar{m} ^2}  
\label{approxdegene}
\eea
\noindent where $\, k_1 (x) = d\,(x h_1) / dx \, ,$ is very small everywhere
except close to the origin, $\bar{m}^2 = (m_R^2+m_L^2 )/2 \, , 
\bar{x} = M_1^2/\bar{m}^2 \, , \bar{y} = |\mu ^2|/\bar{m}^2 \, .$  

\subsection {Special regions in mSUGRA}

One can understand the trend of several of the results presented in
the figures by considering further approximations in the framework
of mSUGRA or similar. Let us first consider the ratio between the
different contributions to the FC component of (\ref{all}) as obtained 
from (\ref{approxI},
\ref{approxI_B}):
\beq
\frac{I_B}{ \frac{1}{2} I_L - I_R}  \approx
\frac{2 \mu^2 \left( I_L - I_R \right) }{ \left( m_R^2 - m_L^2 \right) 
\left( I_L - 2I_R \right) } \approx - \frac{2 \mu^2\, h_1(\bar{x})}
{\bar{m}^2\, g_1(x_R)}
\label{B/1} \eeq
\noindent where $ 1 > h_1 (x) / g_1 (x) > 1/2$  for $ 0 < x < 1 \, .$
From the mSUGRA expression for $\mu^2 $ in terms of $m_R^2$ and $M_1^2 \, ,$
(\ref{mSUGRArelation}), one gets that the $U(1)$ contribution, $I_B + I_L /2 - I_R \, ,$
does not changes sign in the region of physical interest.
The ratio between the $SU(2)$ and the $U(1)$ amplitudes is:  
\beq
\frac{I_2}{ I_B + \frac{1}{2} I_L - I_R}  \approx 
\left( \frac{M_2 \cot^2 \theta_W}{M_1 m_L^2}\right)
\frac{\bar{m}^2\,g_2( x'_L)}{\mu^2\, h_1(\bar{x}) - 
\frac{1}{2} \bar{m}^2\, g_1(\bar{x})}
\label{2/1} 
\eeq
which shows that the chargino term dominates over the neutralino one
as far as $m_R^2 \gg M_1^2 \, .$ Let us now consider the mSUGRA region 
of cosmological interest, $m_R^2 \approx M_1^2 \, , \ m_L^2 \approx
3 M_1^2 \, , \mu^2 \approx 20 M_1^2\, ,$. Therein, one has within the 
approximations given by (\ref{approxI}) and (\ref{approxI_B})
\beq 
I_L \approx 0.17\,{M_1^{-2}}  \qquad  I_R \approx 0.33\,{M_1^{-2}}
\qquad I_B \approx 1.7\,{M_1^{-2}} \qquad 
I_2 \approx 1.05\, M_1^{-2}  \label{cosmoratios} 
\eeq  so that the dominant amplitude is $I_B \, .$ In particular, the
ratio between the $SU(2)$ and the $U(1)$ contributions is $0.72$, as
shown in Fig. (\ref{PL_amusu2u1}). Notice that the $I_B$ dominance in this regime of
mSUGRA is mainly due to the large value of $\mu^2 \, .$  

The analysis of the FV terms is analogous.




\begin{thebibliography}{99}

\bibitem{oscexp} Super-Kamiokande Collaboration,
Phys. Rev. Lett. {\bf 81} (1998) 1562; Phys. Rev. Lett. {\bf 85} (2000) 3999;   
Phys. Rev. Lett. {\bf 86} (2001) 5656, hep-ex/0103033; 
SNO Collaboration, Phys. Rev. Lett. {\bf 87} (2001) 71301, nucl-ex/0106015.

\bibitem{massinsmeth} 
L.J. Hall, V.A. Kostelecky and S. Raby, Nucl. Phys. {\bf B 267} (1986) 415;
F. Gabbiani and A. Masiero, Nucl. Phys. {\bf B 322} (1989) 235.

\bibitem{ggms} 
F. Gabbiani, E. Gabrielli, A. Masiero and L. Silvestrini,  Nucl. Phys. {\bf B 477} (1996) 321, hep-ph/9604387.

\bibitem{cexact} 
J. Hisano, T. Moroi, K. Tobe, M. Yamaguchi and T. Yanagida, Phys. Lett. {\bf B 357} (1995) 579,
hep-ph/9501407; 
J. Hisano, T. Moroi, K. Tobe and M. Yamaguchi, Phys. Rev. {\bf D 53} (1996) 2442, hep-ph/9510309;
J. Hisano and D. Nomura, Phys. Rev. {\b D 59} (1999) 116005, hep-ph/9810479. 

\bibitem{moroi} 
T. Moroi, Phys. Rev. {\bf D 53} (1996) 6565, hep-ph/9512396; Erratum-ibid. {\bf D 56} (1997) 4424.

\bibitem{prs}
S. Pokorski, J. Rosiek and C.A. Savoy, Nucl. Phys. {\bf B 570} (2000) 81, hep-ph/9906206.

\bibitem{fms}
J.L. Feng, K.T. Matchev and Y. Shadmi, Nucl. Phys. {\bf B 613} (2001) 366, hep-ph/0107182.

\bibitem{bormas} F. Borzumati and A. Masiero, Phys. Rev. Lett. {\bf 57} (1986) 961. 

\bibitem{romstr} A. Romanino and A. Strumia, Nucl. Phys. {\bf B 622} (2002) 73, hep-ph/0108275.

\bibitem{deexpp} 
E.D. Commins, S.B. Ross, D. Demille, B.C. Regan,  Phys. Rev. {\bf A 50} (1994) 2960.
\bibitem{deexpf}  B.E. Sauer, talk at {\it Charm, Beauty and CP}, 1st Int. Workshop on Frontier Science,
October 6-11, 2002, Frascati, Italy.  
\bibitem{lam} S.K. Lamoreaux, nucl-ex/0109014. 
\bibitem{dmuexpp} CERN-Mainz-Daresbury Collaboration, Nucl. Phys. {\bf B 150} (1979) 1. 
\bibitem{dmuexpf} R. Carey et al., Letter of Intent to BNL (2000); 
Y.K. Semertzidis et al., hep-ph/0012087.  
\bibitem{dmuexpff} J. Aysto et al., hep-ph/0109217. 
\bibitem{PDB} Particle Data Book, Phys. Rev. {\bf D 66} (2002) 10001.
\bibitem{megexpf} 
L.M. Barkov et al., proposal for an experiment at PSI, http://meg.web.psi.ch.
\bibitem{tmgexpf} 
I. Hinchliffe, F.E. Paige, Phys. Rev. {\bf D 63} (2001) 115006, hep-ph/0010086; 
D.F. Carvalho, J.R. Ellis, M.E. Gomez, S. Lola and J.C. Romao, hep-ph/0206148;
J. Kalinowski, hep-ph/0207051. 


\bibitem{amusus1} 
M. Carena, G.F. Giudice and C.E.M. Wagner, Phys. Lett. {\bf B 390}
(1997) 234, hep-ph/9610233;  
E. Gabrielli and U.  Sarid, Phys. Rev. Lett. {\bf 79} (1997) 4752,
hep-ph/9707546.

\bibitem{amusus2}
J.L. Feng and K.T. Matchev, Phys. Rev. Lett. {\bf 86} (2001) 3480,
hep-ph/0102146;
L. Everett, G.L. Kane, S. Rigolin and L. Wang, Phys.  Rev. Lett. {\bf
86} (2001) 3484, hep-ph/0102145;
T. Ibrahim, U. Chattopadhyay and P. Nath, Phys. Rev. {\bf D 64} (2001)
016010, hep-ph/0102324;
J. Ellis, D.V. Nanopoulos and K.A. Olive, Phys. Lett. {\bf B 508}
(2001) 65, hep-ph/0102331
S. Komine, T. Moroi and M. Yamaguchi, Phys. Lett. {\bf B 507}
(2001) 224, hep-ph/0103182; 
Z. Chacko and G.D. Kribs, Phys. Rev. {\bf D 64} (2001) 75015,
hep-ph/0104317;
D.G. Cerdeno, E. Gabrielli, S. Khalil, C. Munoz and E. Torrente-Lujan,
Phys.  Rev. {\bf D 64} (2001) 093012, hep-ph/0104242;
U. Chattopadhyay and P.  Nath, hep-ph/0208012;
S.P. Martin, J.D. Wells, hep-ph/0209309.

\bibitem{dsus1}
T. Ibrahim and P. Nath, Phys. Rev. {\bf D 58} (1998) 111301, hep-ph/9807501;
T. Falk and K. Olive, Phys. Lett. {\bf B 439} (1998) 71;
M. Brhlik, G. Good and G.L. Kane, Phys. Rev. {\bf D 59} (1999) 115004, hep-ph/9810457;

\bibitem{dsus2}
U. Chattopadhyay, T. Ibrahim and P. Roy, Phys. Rev. {\bf D 64} (2001) 013004, hep-ph/0012337;  
V. Barger et al., hep-ph/0101106;
S. Abel, S. Khalil and O. Lebedev, Nucl. Phys. {\bf B 606} (2001) 151, hep-ph/0103320;
T. Ibrahim and P. Nath, Phys. Rev. {\bf D 64} (2001) 093002, hep-ph/0105025.

\bibitem{cosmo}
J. Ellis, T. Falk and K.A. Olive, Phys. Lett. {\bf B 444} (1998) 367, hep-ph/9810360;
J. Ellis, T. Falk, G. Ganis and K.A. Olive, Phys. Rev. {\bf D 62} (2000) 075010, 
hep-ph/0004169;
J. Ellis, D.V. Nanopoulos and K.A. Olive, Phys. Lett. {\bf B 508} (2001) 65, 
hep-ph/0102331; 
U. Chattopadhyay, A. Corsetti and P. Nath, Phys. Rev. {\bf D 66} (2002) 035003, 
hep-ph/0201001.

\bibitem{kko}   
R. Kitano, M. Koike and Y. Okada, hep-ph/0203110.

\bibitem{meconv}
J. Kaulard et al., Phys. Lett. {\bf B 422} (1998) 334.

\bibitem{meconvf}
M. Bachman et al., (1997) http://meco.ps.uci.edu.    

\bibitem{lms1} S. Lavignac, I. Masina and C.A. Savoy, Phys. Lett. {\bf B 520} (2001) 269, hep-ph/0106245.

\bibitem{ell}
D.F. Carvalho, J. Ellis, M.E. Gomez and S. Lola, Phys. Lett. {\bf B
515} (2001) 323, hep-ph/0103256.

\bibitem{damuexp} G.W. Bennett, et al., Muon g-2 Collaboration, Phys. Rev. Lett. {\bf 89} (2002) 101804; 
Erratum-ibid. 89 (2002) 129903, hep-ex/0208001.

\bibitem{SMpred} M. Davier, S. Eidelman, A. Hocker and Z. Zhang, hep-ph/0208177.

\bibitem{ellis}
J. Ellis, J. Hisano, S. Lola and M. Raidal, Nucl. Phys. {\bf B 621} (2002) 208, hep-ph/0109125;
J. Ellis, J. Hisano, M. Raidal and Y. Shimizu, Phys. Lett. {\bf B 528} (2002) 86, hep-ph/0111324; 
J. Ellis and M. Raidal, Nucl. Phys. {\bf B 643} (2002) 229, hep-ph/0206174. 

\bibitem{imsusy02} I. Masina, hep-ph/0210125.

\bibitem{ms2} I. Masina and C.A. Savoy, in preparation.

\bibitem{bs}     
P. Brax and C.A. Savoy, Nucl. Phys. {\bf B 447} (1995) 227, hep-ph/9503306.

\bibitem{kane}
M. Brhlik, L. Everett, G. L. Kane and J. Lykken, Phys. Rev. {\bf D 62} (2000) 035005, 
hep-ph/9908326; Phys. Rev. Lett. {\bf 83} (1999) 2124, hep-ph/9905215. 


\bibitem{lfvss}

R. Rattazzi and U. Sarid, Nucl. Phys. {\bf B 475} (1996) 27, hep-ph/9512354;
W. Buchmuller, D. Delepine and F. Vissani, Phys. Lett. {\bf B 459} (1999) 171, hep-ph/9904219;
J. L. Feng, Y. Nir and Y. Shadmi, Phys. Rev. {\bf D61} (2000) 113005, hep-ph/9911370;
J. Ellis, M.E. Gomez, G.K. Leontaris, S.Lola and D.V. Nanopoulos, 
Eur. Phys. J. {\bf C 14} (2000) 319, hep-ph/9911459
K.~S.~Babu, B.~Dutta and R.~N.~Mohapatra, Phys.\ Lett.\ {\bf B458} (1999) 93;
W.~Buchmuller, D.~Delepine and L.~T.~Handoko, Nucl.\ Phys.\ {\bf B576} (2000) 445;
J. Sato, K. Tobe and T. Yanagida, Phys. Lett. {\bf B 498} (2001) 189, hep-ph/0010348;
J. Hisano and K. Tobe, Phys. Lett. {\bf B 510} (2001) 197, hep-ph/0102315
J.A. Casas and A. Ibarra, Nucl. Phys. {\bf B 618} (2001) 171, hep-ph/0103065;
S. Davidson and A. Ibarra, JHEP 0109 (2001) 013, hep-ph/0104076; 
T. Blazek and S.F. King, Phys. Lett. {\bf B 518} (2001) 109, hep-ph/0105005;
S. Lavignac, I. Masina and C.A. Savoy, Nucl. Phys. {\bf B 633} (2002) 139, hep-ph/0202086;
A. Masiero, S.K. Vempati and O. Vives, hep-ph/0209303.

\bibitem{barb}
R. Barbieri and L. Hall, Phys. Lett. {\bf B 338} (1994) 212, hep-ph/9408406;
R. Barbieri, L. Hall and A. Strumia, Nucl. Phys. {\bf B 445} (1995) 219, 
hep-ph/9501334; 
R. Barbieri, A. Romanino and A. Strumia, Phys. Lett. {\bf B 369} (1996) 283, 
hep-ph/9511305 
A. Romanino and A. Strumia, Nucl. Phys. {\bf B 490} (1997) 3, hep-ph/9610485. 

\end{thebibliography}
\end{document}